\documentclass[twocolumn,aps,prl,superscriptaddress,preprintnumbers,nofootinbib]{revtex4-2}
\usepackage{graphicx}
\usepackage{subcaption}
\usepackage{color}
\graphicspath{{figures/}{fig/}}
\usepackage{amsmath}
\usepackage{amssymb}
\usepackage{bm}
\usepackage{slashed}
\usepackage{epsfig}
\usepackage{amsfonts}
\usepackage{epstopdf}
\usepackage{hyperref}
\usepackage{bbm}
\usepackage{textcomp}
\usepackage{color}
\usepackage{dsfont}
\usepackage[colorinlistoftodos, textsize=small]{todonotes}
\usepackage{cancel}

\newcommand{\sect}[1]{{\it \textbf{#1.} --- }}
\newcommand{\mi}{\mathrm{i}}

\def\beq{\begin{equation}}
\def\eeq{\end{equation}}
\def\bea{\begin{eqnarray}}
\def\eea{\end{eqnarray}}

\begin{document}
\preprint{SLAC-PUB-251110}

\title{Subleading Color Corrections at Three Loops to the tr$(\phi^2)$ Three-Point Form Factor in $\mathcal{N}$=4 Super Yang-Mills Theory}

\author{Xin Guan}
\email{guanxin@slac.stanford.edu}
\affiliation{SLAC National Accelerator Laboratory, Stanford University, Stanford, CA 94039, USA}

\author{Bernhard Mistlberger}
\email{bernhard.mistlberger@gmail.com}
\affiliation{SLAC National Accelerator Laboratory, Stanford University, Stanford, CA 94039, USA}

\author{Michael Ruf}
\email{mruf@slac.stanford.edu}
\affiliation{SLAC National Accelerator Laboratory, Stanford University, Stanford, CA 94039, USA}

\date{\today}

\begin{abstract}
We compute three-loop corrections to the  three-point form factor of the operator $\operatorname{tr}(\phi^2)$  in $\mathcal{N}=4$ Super Yang-Mills theory. 
In particular, our result is valid beyond the leading-color limit and will consequently be an important input towards extending the amplitude-bootstrap program beyond the leading-color approximation. 
We find that our analytic formulae are strikingly compact expressions in terms of integer linear combinations of generalized polylogarithms of weight six. 
\end{abstract}

\maketitle
\allowdisplaybreaks

\section{Introduction}
Maximally supersymmetric Yang–Mills theory ({$\mathcal{N}=4$ sYM}) provides an ideal laboratory to investigate the properties of four-dimensional Quantum Field Theories.
The study of its observables led to many consequential discoveries, which impact our ability to predict the outcome of real-world scattering events.
Shining examples include the AdS/CFT correspondence~\cite{Maldacena:1997re,Gubser:1998bc,Witten:1998qj}, the Grassmannian geometry~\cite{Arkani-Hamed:2009ljj,Mason:2009qx,Arkani-Hamed:2009nll,Arkani-Hamed:2012zlh}, the amplituhedron~\cite{Arkani-Hamed:2013jha} and the profound connections between scattering amplitudes and Wilson loops~\cite{Drummond:2007aua,Brandhuber:2007yx,Drummond:2007au}. 
Beyond its intrinsic beauty, {$\mathcal{N}=4$ sYM} has served as a fertile testing ground for the modern amplitude developments, inspiring ideas such as unitarity cuts~\cite{Bern:1994zx,Bern:1994cg}, twistor variables~\cite{Witten:2003nn} and symbols~\cite{Goncharov:2010jf}. 
Many of these insights have since been transferred to QCD and gravity, enabling state-of-the-art calculations relevant for LHC phenomenology (see ref.~\cite{Henn:2020omi} for a review) and gravitational-wave physics \cite{Driesse:2024feo,Bern:2025zno}. 
Furthermore, the so-called maximum transcendentality principle~\cite{Kotikov:2002ab,Kotikov:2004er} relates the maximally transcendental part of key QCD quantities --- such as cusp and collinear anomalous dimensions --- to their $\mathcal{N}=4$ sYM counterparts.

The crucial prerequisite to these breakthroughs is the availability of data. 
In this context, data are not in the classical  sense experimentally collected measurements but rather the mathematical predictions of quantities computable from this theory, like scattering amplitudes and form factors.
The explicit computation of the six-particle scattering amplitude in {$\mathcal{N}=4$ sYM}~\cite{DelDuca:2010zg} has led to a breakthrough in our understanding of the mathematical functions spanning scattering amplitudes in any QFT~\cite{Goncharov:2010jf,Duhr:2011zq}.
In turn, understanding how these functions encode the analytic properties of scattering amplitudes and the discovery of new such properties enables more and more powerful computations, i.e. more data.
This virtuous cycle is enshrined in the amplitude bootstrap program.
Currently, this program computed the six-particle amplitude through staggering eight-loop order~\cite{Dixon:2023kop} ---  far beyond our ability to compute with conventional methods --- and led to many other discoveries and results, see for example refs.~\cite{Dixon:2025zwj,Cai:2025atc,Cai:2024znx,Dixon:2022xqh,Dixon:2021tdw,Dixon:2020bbt,Dixon:2020cnr,Caron-Huot:2020bkp,Basso:2020xts,Caron-Huot:2019vjl,Li:2024rkq,Drummond:2014ffa}.

In addition to scattering amplitudes, form factors have drawn considerable attention in recent years; see \cite{Yang:2019vag} for a review. 
Form factors are defined as matrix elements between on-shell asymptotic states and gauge invariant operators.
Consequently, they provide a natural bridge connecting the worlds of amplitudes and correlation functions. 
Of particular importance are the form factors for the stress-tensor multiplet, represented by the operator $\operatorname{tr}(\phi^2)$. 
By virtue of the amplitude bootstrap, the three-point form factor was recently computed through astounding eight-loop order~\cite{Dixon:2020bbt,Dixon:2022rse}. 
The three-loop result has been verified through an explicit analytic calculation in ref.~\cite{Gehrmann:2024tds}.
The four-point form factor is known at two-loops~\cite{Dixon:2024yvq} and form factors with arbitrary number of external legs have been constructed at one-loop \cite{Penante:2014sza}. 
For the  $\operatorname{tr}(\phi^3)$ form factor, results have been derived recently in ref.~\cite{Henn:2024pki,Basso:2024hlx}.

The astonishing progress in  {$\mathcal{N}=4$ sYM} has one limitation -- it is often limited to the so-called planar limit of the theory. 
In this limit, the number of colors $N_\mathrm{c}$ of the $\mathrm{SU}(N_\mathrm{c})$ gauge group is considered to be large and the 't Hooft coupling $a=N_\mathrm{c}g^2/(4\pi)^2$ with the coupling constant $g$ is used as a small expansion parameter. 
In particular, terms suppressed by inverse powers of $N_\mathrm{c}$ (subleading color) are neglected. 
Today, only limited information is available about scattering amplitudes and form factors at subleading color and their properties are far less understood.
Four particle scattering amplitudes at leading color are understood to all-loop order~\cite{Anastasiou:2003kj,Bern:2005iz} while their subleading color contributions are known only through three loops~\cite{Henn:2016jdu}.
Two-loop results are available for five-particle amplitudes~\cite{Chicherin:2018old,Abreu:2018aqd}

In this article, we add an important piece of data to the exploration of {$\mathcal{N}=4$ sYM}  beyond the leading-color limit.
Specifically. we obtain the analytic result for the full-color, three-loop, three-point form factors of the operator tr$(\phi^2)$ in $\mathcal{N}=4$ sYM. 
This constitutes the first full-color analytic result with nontrivial kinematic dependence for a form factor of this operator. 

The  $\operatorname{tr}(\phi^2)$ form factor has particular significance as it is the  {$\mathcal{N}=4$ sYM} equivalent of the scattering amplitude of a Higgs boson and three partons in QCD, after integrating out the degrees of freedom of the top quark.
Consequently, it is linked to some of the most important scattering observables in Higgs boson phenomenology at the Large Hadron Collider. 
One of the most striking observations is that the leading transcendental part of this object in the two theories is identical~\cite{Brandhuber:2012vm}, which was recently confirmed at three loops in ref.~\cite{Chen:2025utl} at three loops in the leading color approximation.
The computation of the subleading-color contributions to the QCD scattering amplitude are still outstanding and the analytic achievements of this article provide a pivotal step towards this goal.

%

\newpage 

\begin{widetext}
\begin{table*}
\begin{center}
\begin{tabular}{ c c c c c c }
\hline\hline
   &    \includegraphics[width=0.15\textwidth]{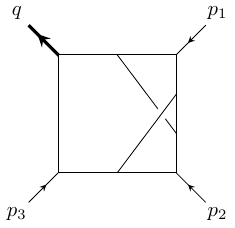} &     \includegraphics[width=0.15\textwidth]{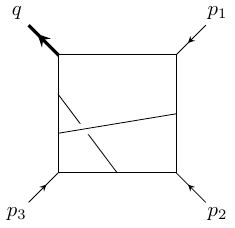}   &  \includegraphics[width=0.15\textwidth]{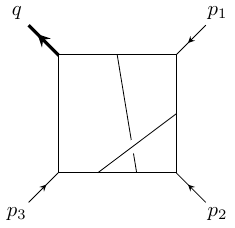} &      \includegraphics[width=0.15\textwidth]{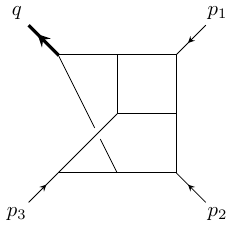} &     \includegraphics[width=0.15\textwidth]{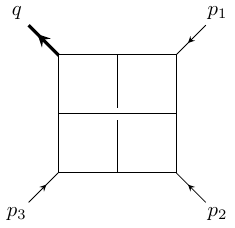} \\
   & (a) & (b) & (c) & (d) & (e)
   \\
  \hline
  Number of MIs & 136 & 174 & 176 & 277 & 371 \\
\hline
  MIs in top sector & 4 & 4 & 4 & 8 & {\bf 19}\\
  \hline
\hline
 
 \vspace{-0.5cm}
\end{tabular}
 \caption{The five new master topologies required for the three-loop three-point form factor of the operator $\operatorname{tr}(\phi^2)$.\label{fig:Integrals}}
\end{center}
 \end{table*}
\end{widetext}


\section{Setup}
\label{sec:set-up}
The three-point form factor for the operator $\mathcal{O}=\operatorname{tr}(\phi^2)$, which is part of the stress-tensor supermultiplet in $\mathcal{N}=4$ sYM, is defined as
\begin{equation}
    \delta^4\left(q{-}{\textstyle\sum}_{i=1}^3p_i\right)\mathcal{F}(p_1,p_2,p_3)=\int\mathrm{d}^4x e^{\mi q\cdot x}\langle 123|\mathcal{O}(x)|0\rangle\,.
\end{equation}
Here $p_i$ are the momenta of three massless states $|i\rangle$, $i\in \{1,2,3\}$ and we have suppressed the color indices associated with these states. 
The kinematics describing this form factor can be parametrized in terms of three ratios of invariants 
\beq
u=\frac{s_{12}}{q^2},\hspace{0.5cm}
v=\frac{s_{23}}{q^2},\hspace{0.5cm}
w=1-u-v=\frac{s_{13}}{q^2}\,,
\eeq
where $s_{ij}=(p_i+p_j)^2$.
The dependence on the remaining invariant $q^2$ is fixed by dimensional analysis and we set $q^2=-1$ for the remainder of the article.
The form factor can be expanded in small t'Hooft coupling
\begin{equation}
    \mathcal{F}=\mathcal{F}^{(0)}\mathcal{I}\,,\quad \mathcal{I}=\sum_{k=0}^\infty a^k \mathcal{I}^{(k)}\,,
\end{equation}
where $\mathcal{F}^{(0)}$ is the leading-order form factor.
Starting at three-loops the form factor receives corrections that are suppressed in the limit of large color representations $N_\mathrm{c}\to \infty$
\begin{equation}
    \mathcal{I}^{(3)}=\mathcal{I}^{(3)}_{\mathrm{LC}}+\frac{1}{N_\mathrm{c}^2}\mathcal{I}^{(3)}_{\mathrm{SLC}}\,.
\end{equation}

\section{Computation and Validation}
The three-loop integrand of our form factor was obtained in ref.~\cite{Lin:2021kht,Lin:2021qol} using color-kinematic duality and unitarity cuts, and was numerically evaluated at one phase point~\cite{Lin:2021qol,Guan:2023gsz}. 
We utilize the package Blade~\cite{Guan:2024byi} to perform Integration-by-parts (IBP) ~\cite{Tkachov1981,Chetyrkin1981,Laporta:2001dd} reduction to express the integrand in terms of so-called master integrals. 
The package Blade incorporates the block-triangular form~\cite{Liu:2018dmc,Guan:2019bcx} method and significantly improves the IBP reduction efficiency.
The residual master integrals are building blocks for any scattering amplitude in QFT with three massless and one massive external particle at third loop order.
A significant fraction of the integrals required for our computation is already publicly available in refs.~\cite{Gehrmann:2024tds,DiVita:2014pza,Canko:2021xmn,Henn:2023vbd,Syrrakos:2023mor}.

For the purposes of this letter, we have to compute an additional set of integrals that can be organized in terms of five integral families depicted in Table~\ref{fig:Integrals}. 
The full set of integrals required for the subleading contribution to the form factor in QCD requires yet a larger set.
However, the missing integrals are less complicated than the most complicated case required here and we defer their calculation to future work.
In order to compute the integrals we use the method of canonical differential equations \cite{Kotikov:1990kg,Kotikov:1991hm,Kotikov:1991pm,Gehrmann:1999as,Henn:2013pwa}, which allows to express the total differential of a set of canonical master integrals $I$ in the following form
\begin{equation}
    \mathrm{d}I=\mathcal{\epsilon}\sum_i A_i\mathrm{d}\log(w_i) I\,, \label{eq:canonicalDE}
\end{equation}
where the $A_i$ are rational number matrices and $\epsilon$ is the dimensional regulator of the space-time dimension $d=4-2\epsilon$.
We constructed the canonical form for the master integrals required for this problem algorithmically using a multivariate version of the algorithm outlined in ref.~\cite{Lee:2014ioa}. 
The application of this algorithm is complicated by the large number of master integrals in the problem. 
For example, the most complicated family depicted in column (e) of tab.~\ref{fig:Integrals} has a total of 371 masters, 19 of which are non-vanishing on the so-called maximal cut.  
We checked that the canonical integrals of refs.~\cite{Gehrmann:2024tds,DiVita:2014pza,Canko:2021xmn,Henn:2023vbd,Syrrakos:2023mor} are rational linear combinations of the canonical basis derived in this work whenever the integrals are shared between the topologies.

We find that the alphabet, the set of letters $w_i$ entering \eqref{eq:canonicalDE}, coincides with the alphabet identified in ref.~\cite{Gehrmann:2024tds}. 
The 20 letters written in terms of the square-root  $r=\sqrt{-uvw}$ are $u,v,w$-permutations of 
\footnote{We count  $20=3+3+6+6+(3-1)$ letters, taking into account the relation \begin{equation}
    \mathrm{d}\log(\tfrac{r+u}{r-u})+\mathrm{d}\log(\tfrac{r+v}{r-v})+\mathrm{d}\log(\tfrac{r+w}{r-w})=0\,.
\end{equation}}
\begin{equation}
    w_i\in\left\{u,1-u,u-(1-v)^2,1-u-uv,\tfrac{r+u}{r-u}\right\}\,.
\end{equation}
Using the canonical differential equation it is straight-forward to evaluate the master integrals to any order in $\epsilon$ in terms of Chen iterated integrals~\cite{Chen:1977oja}. The boundary conditions are determined as in refs.~\cite{Henn:2020lye,Dulat:2014mda,Henn:2013woa} through consistency conditions in the limit $u,v,w\to 0$ \cite{Henn:2020lye} and the evaluation of simple single-scale integrals.
 
Our form factor contains infrared singularities, regulated here by dimensional regularization.
The pattern of this singularities is universal and was predicted in refs.~\cite{Almelid:2015jia,Aybat:2006mz,Aybat:2006wq,Catani:1998bh,Dixon:2008gr,Korchemsky:1987wg,Sterman:2002qn,Becher:2019avh}.
\bea
\label{eq:AFdef}
\mathcal{I}(p_i)&&={\mathbf Z}(\alpha,p_i)\mathcal{I}^{\mathrm{fin}}(p_i),
\eea
where the remainder $\mathcal{I}^{\mathrm{fin}}$ is finite as $d\to 4$. The explicit form of the operator ${\mathbf Z}$ in $\mathcal{N}=4$ sYM is
\begin{align}
&\mathbf{Z}=\exp\Bigg\{ \frac{1}{4} \sum_{L=1}^{\infty} \left(\frac{g^2}{4\pi}\right)^L 
\left[
\frac{\gamma_c^{(L)}}{L^2 \epsilon^2} 
\sum_{i \ne j} \mathbf{T}_i \!\cdot\! \mathbf{T}_j\right.
\\{}&-\left. 
\frac{\gamma_c^{(L)}}{L \epsilon} 
\sum_{i \ne j} \mathbf{T}_i \!\cdot\! \mathbf{T}_j 
\log\!\left( \frac{-s_{ij}}{\mu^2} \right)
+ 
\frac{4}{L \epsilon} \gamma_J^{(L)} \mathbb{I}
+ 
\frac{1}{L \epsilon} \boldsymbol{\Delta}^{(L)}
\right]\Bigg\}\,.\nonumber
\end{align}
Here, $\mu$ is the renormalization scale, $ \mathbf{T}_i$ are the generators of $\mathrm{SU}(N_{\mathrm{c}})$.
The anomalous dimensions $\gamma_c^{(3)}$ and $\gamma_J^{(3)}$ were computed in 
refs.~\cite{Moch:2004pa,Bern:2005iz} and more recently through four loop order in refs.~\cite{Henn:2019swt,vonManteuffel:2020vjv,Agarwal:2021zft}.
%
The infrared singularities of the subleading color part to the form factor exclusively arise from the color-multipole operator $\boldsymbol{\Delta}^{(L)}$. Its contribution to three colored parton amplitudes is given by 
\begin{equation}
\boldsymbol{\Delta}^{(3)}_3 
= -8 c^{(3)}
\sum_{\substack{j<k \\ j,k \ne i}}
f^{abe} f^{cde}
\bigl( 
\mathbf{T}_i^a \mathbf{T}_i^d 
+ 
\mathbf{T}_i^d \mathbf{T}_i^a
\bigr)
\mathbf{T}_j^b \mathbf{T}_k^c \,,
\end{equation}
where $c^{(3)}= \bigl( \zeta_5 + 2 \zeta_2 \zeta_3 \bigr)$ and $f^{abc}$ are the structure constants of $\mathrm{SU}(N_{\mathrm{c}})$.
Given the fact that the subleading color structure first appears at three loops, the divergence is spectacularly mild and can be worked out explicitly as
\begin{equation}
\label{eq:pole}
\mathcal{I}^{(3)}_{\text{SLC}} = -\frac{2c^{(3)}}{\epsilon}+\mathcal{O}(\epsilon^0)\,.
\end{equation}
We use the above formula to extract the finite remainder of the subleading color form factor and study its properties below.

 

We  performed several checks of our result:
\begin{enumerate}
\item We verified that all but the single pole in $\epsilon$ of the subleading color contribution vanish and the remaining pole is correctly predicted by eq.~\eqref{eq:pole}.
\item We checked that our form factor agrees with predictions based on factorization in universal soft~\cite{Herzog:2023sgb,Chen:2023hmk} and collinear limits~\cite{Guan:2024hlf}.
\item  A numeric computation of the $\operatorname{tr}(\phi^2)$ form factor at three loops and subleading color was previously performed for one numerical point refs.~\cite{Lin:2021qol,Guan:2023gsz} and we find agreement with this result.
\end{enumerate}


\section{Analytic structure of the finite remainder}
The finite remainder of the subleading color $\operatorname{tr}(\phi^2)$ form factor exhibits much simpler analytic structure than the master integrals we computed for its evaluation. 
Specifically, we find that it can be written in terms of integer linear combinations of generalized polylogarithms that depend solely on the alphabet
\beq
\mathbb{A}=\label{eq:ffalphabet}
\{u,v,w,1-u,1-v,1-w\},
\eeq
as well as Riemann $\zeta$-values. 
The function is completely symmetric under the permutation of our variables $u$, $v$ and $w$.
This alphabet is identical to the one found for the leading color contributions through eighth loop order~\cite{Dixon:2022rse}.
Next, we study the symbol of the subleading color form factor in analogy to its leading color counterpart, for which a flurry of interesting observations was made. 
The symbol~\cite{Duhr:2011zq,Goncharov:2010jf,Brown:2009qja,Goncharov:2009lql,Chen:1977oja} of the finite part of the form factor takes a very simple form
\begin{equation}
    S(\mathcal{I}^{(3),\mathrm{fin}}_{\mathrm{SLC}})=48\sum_{i=1}^{4506}c_{i}\, w_{i}^{(1)}{\otimes}\dots {\otimes} w_{i}^{(6)}\,,
\end{equation}
where the letters $ w_i^{(k)}$ are drawn from the alphabet $ \mathbb{A}$ in eq.~\eqref{eq:ffalphabet}. The first entry only allows for physical branch cuts $w_{i}^{(1)}\in\{u,v,w\}$.
For all but the last two entries, the symbol satisfies the following adjacency constraint \cite{Dixon:2020bbt,Chicherin:2020umh} and its permutations,
\begin{equation}
    \cancel{\cdots\otimes 1-u\otimes 1-v\otimes\cdots}\,.
\end{equation}
We note that this constraint is only violated by the last two entries of the symbol.
We identify relations among the symbols that are obtained when removing the last entry. 
For example, there is a relation for the total derivative as follows.
\begin{align}
    \text{d} \mathcal{I}^{(3)}_{\mathrm{SLC}}
&={}-12\,c^{(3)}  \text{d}  \log[uvw(1{-}u)(1{-}v)(1{-}w)]\nonumber\\
    &\quad{}+\text{5 independent terms}.
\end{align}
Following the spirit of ref.~\cite{Dixon:2022rse} to find a suitable normalization that removes the above logarithm from our function and that unveils its analytic properties (i.e. a good remainder function), we may choose the normalization
\begin{equation}
    \widetilde{\mathcal{I}} ^{(3)}_{\text{SLC}}
    =\left[uvw(1-u)(1-v)(1-w)\right]^{\frac{1}{2}\epsilon}\mathcal{I} ^{(3)}_{\text{SLC}}\,.
\end{equation}
However, this choice violates the first-entry condition. 
We proceed by analyzing iterative derivatives, i.e. the co-products~\cite{Duhr:2012fh,Brown:2015ylf}, of our function in order to find further possible hidden constraints. 
The number of independent functions in the co-product gives an indication of this and we tabulate these numbers in Table~\ref{tab:CoproductSize}.
We find non-trivial relations among these co-products (beyond the so-called integrability constraint).
In particular, the totally symmetric combination $[uvw(1-u)(1-v)(1-w)]$ can never appear in the last entry of $\widetilde{\mathcal{I}} ^{(3)}_{\text{SLC}}$.
\begin{table}
\begin{tabular}{lccccccc}
\hline
     $n$&0&1&2&3&4&5&6  \\
     \hline
     subleading color &1&3&10&27&21&5&1  \\
     leading color \cite{Dixon:2020bbt} &1 &3 &9 &12&6& 3&1\\
     \hline
\end{tabular}
\caption{Number of independent functions in the co-product $\Delta_{n,1,\dots 1}$ of the three loop remainder $\widetilde{\mathcal{I}} ^{(3)}_{\text{SLC}}$.
\label{tab:CoproductSize}
}
\end{table}

\begin{table}
\begin{tabular}{l c c c c c c c}
\hline
loops     & $1$  & $2$ & $ \hspace{0,5 cm}3-\text{LC}$  \hspace{0,5 cm} &$3-\text{SLC}$ \\
      \hline
$(u,\infty,-\infty)$ HPLs &  1  & 2 & 14 & 0 \\
$(1,v,-v)$ HPLs &  2 & 4 & 54 & 78 \\
$(u,u,1-2u)$ HPLs  \hspace{0,5 cm} &  4 & 25 & 269 & 291 \\
     \hline
\end{tabular}
\caption{Number of terms in the symbols at $n$-loops at three kinematic lines of the variables $(u,v,w)$.
\label{tab:LineSize}
}
\end{table}
We find a space of 118 possible symbols satisfying integrability, $u,v,w$ permutation invariance, the first entry condition, the modified adjacency constraint and the last entry condition and match the predicted behaviour in the collinear limit.
It was observed in ref.~\cite{Dixon:2020bbt} that the space of functions in the form factor collapses to so-called harmonic polylogarithms (HPLs)~\cite{Remiddi:1999ew} on three kinematic lines $(u,\infty,-\infty)$, $(1,v,-v)$ and $(u,u,1-2u)$ of the variables $(u,v,w)$. 

In the limit $v\to \infty$, we find at fixed $u$,
{\begin{align}
&\mathcal{I}^{(3),\mathrm{fin}}_{\mathrm{SLC}}={}\nonumber\\*
&\pi^{2} \Big[
 96 \big( H_{0,0,1,0}(u) {+} H_{0,1,0,0}(u) {-} H_{1,0,1,0}(u) {-} H_{1,1,0,0}(u) \big)
\nonumber\\*
&\quad {+} 24 \zeta_{3} (\log(u){+}4 \log(1 {-} u){-}4\log(v){+}2\mi\pi)\, 
{-} \frac{1406\pi^{4}}{945} \Bigg]\nonumber\\*
&\quad -24 \big( \zeta_{3}^{2} - 3 \log(u)\, \zeta_{5} \big){+}\mathcal{O}\left(\frac{1}{v}\right)\label{eq:functionLargev}\,,
\end{align}}
where $H$ denotes a harmonic polylogarithm.
Several structures in this expression, in particular the single-logarithmic divergence and the $\pi^2$ are suggestive of a high energy factorization. Deriving this structure from an effective field theory, provides an interesting task for the future, but is beyond the scope of this letter.

More generally, we choose to parametrize the special kinematic lines using the variables $1-1/u$, $1/v$ and $u$ respectively and compute the symbol of our function. 
Table~\ref{tab:LineSize} reports the number of terms in the symbol on these three lines. 
Leading color results agree with ref.~\cite{Dixon:2020bbt}.
Note, that the symbol of the function \eqref{eq:functionLargev} vanishes for the subleading color form factor on the line $(u,\infty,-\infty)$.
Furthermore, we evaluated our result at the symmetric point $u=v=w=1/3,\, q^2=-6$. 
We used the tool \texttt{HyperlogProcedures}~\cite{Schnetz2025_HyperlogProcedures} to convert all the HPLs of argument $1/3$ into cyclotomic zeta values. 
Our result reproduces the numeric results of ref.~\cite{Guan:2023gsz} to  30 digits and is attached to the arXiv submission of this article. 


\begin{figure}
    \centering
    \includegraphics[width=0.9\linewidth]{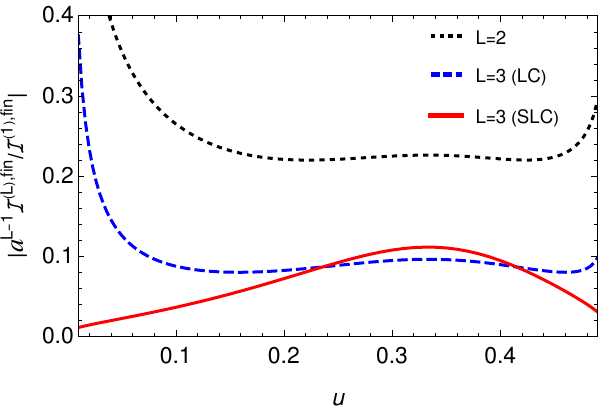}
   \caption{The absolute value of the ratio of the corrections of the finite remainder function at L-loops to its 1-loop counterpart in the Euclidean region.}
    \label{fig:finite_remainder}
\end{figure}
Finally, we study the numerical impact of the subleading color contributions to the form factor evaluating the function using publicly available tools~\cite{Bauer2000,zhenjiempl}.  
Specifically, we look at the ratio of different loop-order contributions to the one-loop correction in order to asses their relative perturbative importance.
Fig~\ref{fig:finite_remainder} shows the absolute value of the ratio along the line $v=u$ in the Euclidean region for $q^2=-1$. 
The different orders include the value of the coupling constant $a=0.118 \times 3/(4\pi)$ and we set $N_\mathrm{c}=3$. 
The subleading color contribution to the form factor in $\mathcal{N}=4$ might serve as a proxy to estimate the numerical importance of subleading color contributions to form factors in other gauge theories and might consequently indicate their importance to Higgs boson physics and QCD. 
We observe in fig.~\ref{fig:finite_remainder} that the subleading color contributions are indeed of comparable size to their leading order three-loop counter-part. 
Furthermore, the shape of the subleading color contribution is quite different, owing largely to the different logarithmic structure at the boundaries of the plot.



\section{Conclusions}
\label{sec:conclusions}

In this letter we presented the first analytic computation of a three-point form factor in four-dimensional gauge theory at third loop order, including subleading color contributions.
Specifically, we computed the $\operatorname{tr}(\phi^2)$ form factor in $\mathcal{N}=4$ sYM theory through three loops. 
Our analytic results are attached to the arXiv submission of this article in electronically readable files.

The analytic structure of the subleading color contribution to the form factor are remarkably simple. 
They can be expressed in terms of an integer linear combination of the same class of generalized polylogarithms that appear in the leading color contribution. 
This simplicity is highly non-trivial and arises out of intricate cancellations of the much more complex functions of the master integrals we computed to obtain this result.
Furthermore, we study the symbol of the newly computed form factor and discover intriguing relations that may hint at additional symmetry and simplicity of this object beyond the leading color limit.

The simplicity of the analytic structure of the leading color form factor has enabled the amplitude bootstrap of this quantity to record setting eight-loop order. 
Currently, an extension to subleading color of this approach does not exist and our result provides a first step toward such a subleading color form-factor bootstrap. 
In particular, one might speculate that the presence of a weakened versions of the last-entry and adjacency constraints~\cite{Caron-Huot:2019bsq,Drummond:2017ssj} could enable such an approach.

Our calculation required evaluating five previously unknown families of three-loop master integrals with one off-shell leg representing a major step toward the full-color computation of  scattering amplitudes for a Higgs boson or gauge boson and three partons.
The integrals obtained for this letter include the most complex topologies needed for this task.  
We observe that subleading color effects can be sizable, further motivating their calculation in QCD, which we leave to future work.



\begin{acknowledgments}
	\sect{Acknowledgments}
	We thank Lance Dixon for useful discussions and comments on the draft and Michael Saavedra for useful discussions. XG, BM, MR are supported by the United States Department of Energy, Contract DE-AC02-76SF00515. 
\end{acknowledgments}

\bibliography{refs.bib}

\begin{thebibliography}{98}%
\makeatletter
\providecommand \@ifxundefined [1]{%
 \@ifx{#1\undefined}
}%
\providecommand \@ifnum [1]{%
 \ifnum #1\expandafter \@firstoftwo
 \else \expandafter \@secondoftwo
 \fi
}%
\providecommand \@ifx [1]{%
 \ifx #1\expandafter \@firstoftwo
 \else \expandafter \@secondoftwo
 \fi
}%
\providecommand \natexlab [1]{#1}%
\providecommand \enquote  [1]{``#1''}%
\providecommand \bibnamefont  [1]{#1}%
\providecommand \bibfnamefont [1]{#1}%
\providecommand \citenamefont [1]{#1}%
\providecommand \href@noop [0]{\@secondoftwo}%
\providecommand \href [0]{\begingroup \@sanitize@url \@href}%
\providecommand \@href[1]{\@@startlink{#1}\@@href}%
\providecommand \@@href[1]{\endgroup#1\@@endlink}%
\providecommand \@sanitize@url [0]{\catcode `\\12\catcode `\$12\catcode
  `\&12\catcode `\#12\catcode `\^12\catcode `\_12\catcode `\%12\relax}%
\providecommand \@@startlink[1]{}%
\providecommand \@@endlink[0]{}%
\providecommand \url  [0]{\begingroup\@sanitize@url \@url }%
\providecommand \@url [1]{\endgroup\@href {#1}{\urlprefix }}%
\providecommand \urlprefix  [0]{URL }%
\providecommand \Eprint [0]{\href }%
\providecommand \doibase [0]{https://doi.org/}%
\providecommand \selectlanguage [0]{\@gobble}%
\providecommand \bibinfo  [0]{\@secondoftwo}%
\providecommand \bibfield  [0]{\@secondoftwo}%
\providecommand \translation [1]{[#1]}%
\providecommand \BibitemOpen [0]{}%
\providecommand \bibitemStop [0]{}%
\providecommand \bibitemNoStop [0]{.\EOS\space}%
\providecommand \EOS [0]{\spacefactor3000\relax}%
\providecommand \BibitemShut  [1]{\csname bibitem#1\endcsname}%
\let\auto@bib@innerbib\@empty
\bibitem [{\citenamefont {Maldacena}(1998)}]{Maldacena:1997re}%
  \BibitemOpen
  \bibfield  {author} {\bibinfo {author} {\bibfnamefont {J.~M.}\ \bibnamefont
  {Maldacena}},\ }\bibfield  {title} {\bibinfo {title} {{The Large $N$ limit of
  superconformal field theories and supergravity}},\ }\href
  {https://doi.org/10.4310/ATMP.1998.v2.n2.a1} {\bibfield  {journal} {\bibinfo
  {journal} {Adv. Theor. Math. Phys.}\ }\textbf {\bibinfo {volume} {2}},\
  \bibinfo {pages} {231} (\bibinfo {year} {1998})},\ \Eprint
  {https://arxiv.org/abs/hep-th/9711200} {arXiv:hep-th/9711200} \BibitemShut
  {NoStop}%
\bibitem [{\citenamefont {Gubser}\ \emph {et~al.}(1998)\citenamefont {Gubser},
  \citenamefont {Klebanov},\ and\ \citenamefont {Polyakov}}]{Gubser:1998bc}%
  \BibitemOpen
  \bibfield  {author} {\bibinfo {author} {\bibfnamefont {S.~S.}\ \bibnamefont
  {Gubser}}, \bibinfo {author} {\bibfnamefont {I.~R.}\ \bibnamefont
  {Klebanov}},\ and\ \bibinfo {author} {\bibfnamefont {A.~M.}\ \bibnamefont
  {Polyakov}},\ }\bibfield  {title} {\bibinfo {title} {{Gauge theory
  correlators from noncritical string theory}},\ }\href
  {https://doi.org/10.1016/S0370-2693(98)00377-3} {\bibfield  {journal}
  {\bibinfo  {journal} {Phys. Lett. B}\ }\textbf {\bibinfo {volume} {428}},\
  \bibinfo {pages} {105} (\bibinfo {year} {1998})},\ \Eprint
  {https://arxiv.org/abs/hep-th/9802109} {arXiv:hep-th/9802109} \BibitemShut
  {NoStop}%
\bibitem [{\citenamefont {Witten}(1998)}]{Witten:1998qj}%
  \BibitemOpen
  \bibfield  {author} {\bibinfo {author} {\bibfnamefont {E.}~\bibnamefont
  {Witten}},\ }\bibfield  {title} {\bibinfo {title} {{Anti de Sitter space and
  holography}},\ }\href {https://doi.org/10.4310/ATMP.1998.v2.n2.a2} {\bibfield
   {journal} {\bibinfo  {journal} {Adv. Theor. Math. Phys.}\ }\textbf {\bibinfo
  {volume} {2}},\ \bibinfo {pages} {253} (\bibinfo {year} {1998})},\ \Eprint
  {https://arxiv.org/abs/hep-th/9802150} {arXiv:hep-th/9802150} \BibitemShut
  {NoStop}%
\bibitem [{\citenamefont {Arkani-Hamed}\ \emph
  {et~al.}(2010{\natexlab{a}})\citenamefont {Arkani-Hamed}, \citenamefont
  {Cachazo}, \citenamefont {Cheung},\ and\ \citenamefont
  {Kaplan}}]{Arkani-Hamed:2009ljj}%
  \BibitemOpen
  \bibfield  {author} {\bibinfo {author} {\bibfnamefont {N.}~\bibnamefont
  {Arkani-Hamed}}, \bibinfo {author} {\bibfnamefont {F.}~\bibnamefont
  {Cachazo}}, \bibinfo {author} {\bibfnamefont {C.}~\bibnamefont {Cheung}},\
  and\ \bibinfo {author} {\bibfnamefont {J.}~\bibnamefont {Kaplan}},\
  }\bibfield  {title} {\bibinfo {title} {{A Duality For The S Matrix}},\ }\href
  {https://doi.org/10.1007/JHEP03(2010)020} {\bibfield  {journal} {\bibinfo
  {journal} {JHEP}\ }\textbf {\bibinfo {volume} {03}},\ \bibinfo {pages}
  {020}},\ \Eprint {https://arxiv.org/abs/0907.5418} {arXiv:0907.5418 [hep-th]}
  \BibitemShut {NoStop}%
\bibitem [{\citenamefont {Mason}\ and\ \citenamefont
  {Skinner}(2009)}]{Mason:2009qx}%
  \BibitemOpen
  \bibfield  {author} {\bibinfo {author} {\bibfnamefont {L.~J.}\ \bibnamefont
  {Mason}}\ and\ \bibinfo {author} {\bibfnamefont {D.}~\bibnamefont
  {Skinner}},\ }\bibfield  {title} {\bibinfo {title} {{Dual Superconformal
  Invariance, Momentum Twistors and Grassmannians}},\ }\href
  {https://doi.org/10.1088/1126-6708/2009/11/045} {\bibfield  {journal}
  {\bibinfo  {journal} {JHEP}\ }\textbf {\bibinfo {volume} {11}},\ \bibinfo
  {pages} {045}},\ \Eprint {https://arxiv.org/abs/0909.0250} {arXiv:0909.0250
  [hep-th]} \BibitemShut {NoStop}%
\bibitem [{\citenamefont {Arkani-Hamed}\ \emph
  {et~al.}(2010{\natexlab{b}})\citenamefont {Arkani-Hamed}, \citenamefont
  {Cachazo},\ and\ \citenamefont {Cheung}}]{Arkani-Hamed:2009nll}%
  \BibitemOpen
  \bibfield  {author} {\bibinfo {author} {\bibfnamefont {N.}~\bibnamefont
  {Arkani-Hamed}}, \bibinfo {author} {\bibfnamefont {F.}~\bibnamefont
  {Cachazo}},\ and\ \bibinfo {author} {\bibfnamefont {C.}~\bibnamefont
  {Cheung}},\ }\bibfield  {title} {\bibinfo {title} {{The Grassmannian Origin
  Of Dual Superconformal Invariance}},\ }\href
  {https://doi.org/10.1007/JHEP03(2010)036} {\bibfield  {journal} {\bibinfo
  {journal} {JHEP}\ }\textbf {\bibinfo {volume} {03}},\ \bibinfo {pages}
  {036}},\ \Eprint {https://arxiv.org/abs/0909.0483} {arXiv:0909.0483 [hep-th]}
  \BibitemShut {NoStop}%
\bibitem [{\citenamefont {Arkani-Hamed}\ \emph {et~al.}(2016)\citenamefont
  {Arkani-Hamed}, \citenamefont {Bourjaily}, \citenamefont {Cachazo},
  \citenamefont {Goncharov}, \citenamefont {Postnikov},\ and\ \citenamefont
  {Trnka}}]{Arkani-Hamed:2012zlh}%
  \BibitemOpen
  \bibfield  {author} {\bibinfo {author} {\bibfnamefont {N.}~\bibnamefont
  {Arkani-Hamed}}, \bibinfo {author} {\bibfnamefont {J.~L.}\ \bibnamefont
  {Bourjaily}}, \bibinfo {author} {\bibfnamefont {F.}~\bibnamefont {Cachazo}},
  \bibinfo {author} {\bibfnamefont {A.~B.}\ \bibnamefont {Goncharov}}, \bibinfo
  {author} {\bibfnamefont {A.}~\bibnamefont {Postnikov}},\ and\ \bibinfo
  {author} {\bibfnamefont {J.}~\bibnamefont {Trnka}},\ }\href
  {https://doi.org/10.1017/CBO9781316091548} {\emph {\bibinfo {title}
  {{Grassmannian Geometry of Scattering Amplitudes}}}}\ (\bibinfo  {publisher}
  {Cambridge University Press},\ \bibinfo {year} {2016})\ \Eprint
  {https://arxiv.org/abs/1212.5605} {arXiv:1212.5605 [hep-th]} \BibitemShut
  {NoStop}%
\bibitem [{\citenamefont {Arkani-Hamed}\ and\ \citenamefont
  {Trnka}(2014)}]{Arkani-Hamed:2013jha}%
  \BibitemOpen
  \bibfield  {author} {\bibinfo {author} {\bibfnamefont {N.}~\bibnamefont
  {Arkani-Hamed}}\ and\ \bibinfo {author} {\bibfnamefont {J.}~\bibnamefont
  {Trnka}},\ }\bibfield  {title} {\bibinfo {title} {{The Amplituhedron}},\
  }\href {https://doi.org/10.1007/JHEP10(2014)030} {\bibfield  {journal}
  {\bibinfo  {journal} {JHEP}\ }\textbf {\bibinfo {volume} {10}},\ \bibinfo
  {pages} {030}},\ \Eprint {https://arxiv.org/abs/1312.2007} {arXiv:1312.2007
  [hep-th]} \BibitemShut {NoStop}%
\bibitem [{\citenamefont {Drummond}\ \emph {et~al.}(2008)\citenamefont
  {Drummond}, \citenamefont {Korchemsky},\ and\ \citenamefont
  {Sokatchev}}]{Drummond:2007aua}%
  \BibitemOpen
  \bibfield  {author} {\bibinfo {author} {\bibfnamefont {J.~M.}\ \bibnamefont
  {Drummond}}, \bibinfo {author} {\bibfnamefont {G.~P.}\ \bibnamefont
  {Korchemsky}},\ and\ \bibinfo {author} {\bibfnamefont {E.}~\bibnamefont
  {Sokatchev}},\ }\bibfield  {title} {\bibinfo {title} {{Conformal properties
  of four-gluon planar amplitudes and Wilson loops}},\ }\href
  {https://doi.org/10.1016/j.nuclphysb.2007.11.041} {\bibfield  {journal}
  {\bibinfo  {journal} {Nucl. Phys. B}\ }\textbf {\bibinfo {volume} {795}},\
  \bibinfo {pages} {385} (\bibinfo {year} {2008})},\ \Eprint
  {https://arxiv.org/abs/0707.0243} {arXiv:0707.0243 [hep-th]} \BibitemShut
  {NoStop}%
\bibitem [{\citenamefont {Brandhuber}\ \emph {et~al.}(2008)\citenamefont
  {Brandhuber}, \citenamefont {Heslop},\ and\ \citenamefont
  {Travaglini}}]{Brandhuber:2007yx}%
  \BibitemOpen
  \bibfield  {author} {\bibinfo {author} {\bibfnamefont {A.}~\bibnamefont
  {Brandhuber}}, \bibinfo {author} {\bibfnamefont {P.}~\bibnamefont {Heslop}},\
  and\ \bibinfo {author} {\bibfnamefont {G.}~\bibnamefont {Travaglini}},\
  }\bibfield  {title} {\bibinfo {title} {{MHV amplitudes in N=4 super
  Yang-Mills and Wilson loops}},\ }\href
  {https://doi.org/10.1016/j.nuclphysb.2007.11.002} {\bibfield  {journal}
  {\bibinfo  {journal} {Nucl. Phys. B}\ }\textbf {\bibinfo {volume} {794}},\
  \bibinfo {pages} {231} (\bibinfo {year} {2008})},\ \Eprint
  {https://arxiv.org/abs/0707.1153} {arXiv:0707.1153 [hep-th]} \BibitemShut
  {NoStop}%
\bibitem [{\citenamefont {Drummond}\ \emph {et~al.}(2010)\citenamefont
  {Drummond}, \citenamefont {Henn}, \citenamefont {Korchemsky},\ and\
  \citenamefont {Sokatchev}}]{Drummond:2007au}%
  \BibitemOpen
  \bibfield  {author} {\bibinfo {author} {\bibfnamefont {J.~M.}\ \bibnamefont
  {Drummond}}, \bibinfo {author} {\bibfnamefont {J.}~\bibnamefont {Henn}},
  \bibinfo {author} {\bibfnamefont {G.~P.}\ \bibnamefont {Korchemsky}},\ and\
  \bibinfo {author} {\bibfnamefont {E.}~\bibnamefont {Sokatchev}},\ }\bibfield
  {title} {\bibinfo {title} {{Conformal Ward identities for Wilson loops and a
  test of the duality with gluon amplitudes}},\ }\href
  {https://doi.org/10.1016/j.nuclphysb.2009.10.013} {\bibfield  {journal}
  {\bibinfo  {journal} {Nucl. Phys. B}\ }\textbf {\bibinfo {volume} {826}},\
  \bibinfo {pages} {337} (\bibinfo {year} {2010})},\ \Eprint
  {https://arxiv.org/abs/0712.1223} {arXiv:0712.1223 [hep-th]} \BibitemShut
  {NoStop}%
\bibitem [{\citenamefont {Bern}\ \emph {et~al.}(1994)\citenamefont {Bern},
  \citenamefont {Dixon}, \citenamefont {Dunbar},\ and\ \citenamefont
  {Kosower}}]{Bern:1994zx}%
  \BibitemOpen
  \bibfield  {author} {\bibinfo {author} {\bibfnamefont {Z.}~\bibnamefont
  {Bern}}, \bibinfo {author} {\bibfnamefont {L.~J.}\ \bibnamefont {Dixon}},
  \bibinfo {author} {\bibfnamefont {D.~C.}\ \bibnamefont {Dunbar}},\ and\
  \bibinfo {author} {\bibfnamefont {D.~A.}\ \bibnamefont {Kosower}},\
  }\bibfield  {title} {\bibinfo {title} {{One loop n point gauge theory
  amplitudes, unitarity and collinear limits}},\ }\href
  {https://doi.org/10.1016/0550-3213(94)90179-1} {\bibfield  {journal}
  {\bibinfo  {journal} {Nucl. Phys. B}\ }\textbf {\bibinfo {volume} {425}},\
  \bibinfo {pages} {217} (\bibinfo {year} {1994})},\ \Eprint
  {https://arxiv.org/abs/hep-ph/9403226} {arXiv:hep-ph/9403226} \BibitemShut
  {NoStop}%
\bibitem [{\citenamefont {Bern}\ \emph {et~al.}(1995)\citenamefont {Bern},
  \citenamefont {Dixon}, \citenamefont {Dunbar},\ and\ \citenamefont
  {Kosower}}]{Bern:1994cg}%
  \BibitemOpen
  \bibfield  {author} {\bibinfo {author} {\bibfnamefont {Z.}~\bibnamefont
  {Bern}}, \bibinfo {author} {\bibfnamefont {L.~J.}\ \bibnamefont {Dixon}},
  \bibinfo {author} {\bibfnamefont {D.~C.}\ \bibnamefont {Dunbar}},\ and\
  \bibinfo {author} {\bibfnamefont {D.~A.}\ \bibnamefont {Kosower}},\
  }\bibfield  {title} {\bibinfo {title} {{Fusing gauge theory tree amplitudes
  into loop amplitudes}},\ }\href
  {https://doi.org/10.1016/0550-3213(94)00488-Z} {\bibfield  {journal}
  {\bibinfo  {journal} {Nucl. Phys. B}\ }\textbf {\bibinfo {volume} {435}},\
  \bibinfo {pages} {59} (\bibinfo {year} {1995})},\ \Eprint
  {https://arxiv.org/abs/hep-ph/9409265} {arXiv:hep-ph/9409265} \BibitemShut
  {NoStop}%
\bibitem [{\citenamefont {Witten}(2004)}]{Witten:2003nn}%
  \BibitemOpen
  \bibfield  {author} {\bibinfo {author} {\bibfnamefont {E.}~\bibnamefont
  {Witten}},\ }\bibfield  {title} {\bibinfo {title} {{Perturbative gauge theory
  as a string theory in twistor space}},\ }\href
  {https://doi.org/10.1007/s00220-004-1187-3} {\bibfield  {journal} {\bibinfo
  {journal} {Commun. Math. Phys.}\ }\textbf {\bibinfo {volume} {252}},\
  \bibinfo {pages} {189} (\bibinfo {year} {2004})},\ \Eprint
  {https://arxiv.org/abs/hep-th/0312171} {arXiv:hep-th/0312171} \BibitemShut
  {NoStop}%
\bibitem [{\citenamefont {Goncharov}\ \emph {et~al.}(2010)\citenamefont
  {Goncharov}, \citenamefont {Spradlin}, \citenamefont {Vergu},\ and\
  \citenamefont {Volovich}}]{Goncharov:2010jf}%
  \BibitemOpen
  \bibfield  {author} {\bibinfo {author} {\bibfnamefont {A.~B.}\ \bibnamefont
  {Goncharov}}, \bibinfo {author} {\bibfnamefont {M.}~\bibnamefont {Spradlin}},
  \bibinfo {author} {\bibfnamefont {C.}~\bibnamefont {Vergu}},\ and\ \bibinfo
  {author} {\bibfnamefont {A.}~\bibnamefont {Volovich}},\ }\bibfield  {title}
  {\bibinfo {title} {{Classical Polylogarithms for Amplitudes and Wilson
  Loops}},\ }\href {https://doi.org/10.1103/PhysRevLett.105.151605} {\bibfield
  {journal} {\bibinfo  {journal} {Phys. Rev. Lett.}\ }\textbf {\bibinfo
  {volume} {105}},\ \bibinfo {pages} {151605} (\bibinfo {year} {2010})},\
  \Eprint {https://arxiv.org/abs/1006.5703} {arXiv:1006.5703 [hep-th]}
  \BibitemShut {NoStop}%
\bibitem [{\citenamefont {Henn}(2021)}]{Henn:2020omi}%
  \BibitemOpen
  \bibfield  {author} {\bibinfo {author} {\bibfnamefont {J.~M.}\ \bibnamefont
  {Henn}},\ }\bibfield  {title} {\bibinfo {title} {{What Can We Learn About QCD
  and Collider Physics from N=4 Super Yang{\textendash}Mills?}},\ }\href
  {https://doi.org/10.1146/annurev-nucl-102819-100428} {\bibfield  {journal}
  {\bibinfo  {journal} {Ann. Rev. Nucl. Part. Sci.}\ }\textbf {\bibinfo
  {volume} {71}},\ \bibinfo {pages} {87} (\bibinfo {year} {2021})},\ \Eprint
  {https://arxiv.org/abs/2006.00361} {arXiv:2006.00361 [hep-th]} \BibitemShut
  {NoStop}%
\bibitem [{\citenamefont {Driesse}\ \emph {et~al.}(2025)\citenamefont
  {Driesse}, \citenamefont {Jakobsen}, \citenamefont {Klemm}, \citenamefont
  {Mogull}, \citenamefont {Nega}, \citenamefont {Plefka}, \citenamefont
  {Sauer},\ and\ \citenamefont {Usovitsch}}]{Driesse:2024feo}%
  \BibitemOpen
  \bibfield  {author} {\bibinfo {author} {\bibfnamefont {M.}~\bibnamefont
  {Driesse}}, \bibinfo {author} {\bibfnamefont {G.~U.}\ \bibnamefont
  {Jakobsen}}, \bibinfo {author} {\bibfnamefont {A.}~\bibnamefont {Klemm}},
  \bibinfo {author} {\bibfnamefont {G.}~\bibnamefont {Mogull}}, \bibinfo
  {author} {\bibfnamefont {C.}~\bibnamefont {Nega}}, \bibinfo {author}
  {\bibfnamefont {J.}~\bibnamefont {Plefka}}, \bibinfo {author} {\bibfnamefont
  {B.}~\bibnamefont {Sauer}},\ and\ \bibinfo {author} {\bibfnamefont
  {J.}~\bibnamefont {Usovitsch}},\ }\bibfield  {title} {\bibinfo {title}
  {{Emergence of Calabi{\textendash}Yau manifolds in high-precision black-hole
  scattering}},\ }\href {https://doi.org/10.1038/s41586-025-08984-2} {\bibfield
   {journal} {\bibinfo  {journal} {Nature}\ }\textbf {\bibinfo {volume}
  {641}},\ \bibinfo {pages} {603} (\bibinfo {year} {2025})},\ \Eprint
  {https://arxiv.org/abs/2411.11846} {arXiv:2411.11846 [hep-th]} \BibitemShut
  {NoStop}%
\bibitem [{\citenamefont {Bern}\ \emph {et~al.}(2025)\citenamefont {Bern},
  \citenamefont {Herrmann}, \citenamefont {Roiban}, \citenamefont {Ruf},
  \citenamefont {Smirnov}, \citenamefont {Smirnov},\ and\ \citenamefont
  {Zeng}}]{Bern:2025zno}%
  \BibitemOpen
  \bibfield  {author} {\bibinfo {author} {\bibfnamefont {Z.}~\bibnamefont
  {Bern}}, \bibinfo {author} {\bibfnamefont {E.}~\bibnamefont {Herrmann}},
  \bibinfo {author} {\bibfnamefont {R.}~\bibnamefont {Roiban}}, \bibinfo
  {author} {\bibfnamefont {M.~S.}\ \bibnamefont {Ruf}}, \bibinfo {author}
  {\bibfnamefont {A.~V.}\ \bibnamefont {Smirnov}}, \bibinfo {author}
  {\bibfnamefont {V.~A.}\ \bibnamefont {Smirnov}},\ and\ \bibinfo {author}
  {\bibfnamefont {M.}~\bibnamefont {Zeng}},\ }\bibfield  {title} {\bibinfo
  {title} {{Second-order self-force potential-region binary dynamics at
  $O(G^5)$ in supergravity}},\ }\href@noop {} {\  (\bibinfo {year} {2025})},\
  \Eprint {https://arxiv.org/abs/2509.17412} {arXiv:2509.17412 [hep-th]}
  \BibitemShut {NoStop}%
\bibitem [{\citenamefont {Kotikov}\ and\ \citenamefont
  {Lipatov}(2003)}]{Kotikov:2002ab}%
  \BibitemOpen
  \bibfield  {author} {\bibinfo {author} {\bibfnamefont {A.~V.}\ \bibnamefont
  {Kotikov}}\ and\ \bibinfo {author} {\bibfnamefont {L.~N.}\ \bibnamefont
  {Lipatov}},\ }\bibfield  {title} {\bibinfo {title} {{DGLAP and BFKL equations
  in the $N=4$ supersymmetric gauge theory}},\ }\href
  {https://doi.org/10.1016/S0550-3213(03)00264-5} {\bibfield  {journal}
  {\bibinfo  {journal} {Nucl. Phys. B}\ }\textbf {\bibinfo {volume} {661}},\
  \bibinfo {pages} {19} (\bibinfo {year} {2003})},\ \bibinfo {note} {[Erratum:
  Nucl.Phys.B 685, 405--407 (2004)]},\ \Eprint
  {https://arxiv.org/abs/hep-ph/0208220} {arXiv:hep-ph/0208220} \BibitemShut
  {NoStop}%
\bibitem [{\citenamefont {Kotikov}\ \emph {et~al.}(2004)\citenamefont
  {Kotikov}, \citenamefont {Lipatov}, \citenamefont {Onishchenko},\ and\
  \citenamefont {Velizhanin}}]{Kotikov:2004er}%
  \BibitemOpen
  \bibfield  {author} {\bibinfo {author} {\bibfnamefont {A.~V.}\ \bibnamefont
  {Kotikov}}, \bibinfo {author} {\bibfnamefont {L.~N.}\ \bibnamefont
  {Lipatov}}, \bibinfo {author} {\bibfnamefont {A.~I.}\ \bibnamefont
  {Onishchenko}},\ and\ \bibinfo {author} {\bibfnamefont {V.~N.}\ \bibnamefont
  {Velizhanin}},\ }\bibfield  {title} {\bibinfo {title} {{Three loop universal
  anomalous dimension of the Wilson operators in $N=4$ SUSY Yang-Mills
  model}},\ }\href {https://doi.org/10.1016/j.physletb.2004.05.078} {\bibfield
  {journal} {\bibinfo  {journal} {Phys. Lett. B}\ }\textbf {\bibinfo {volume}
  {595}},\ \bibinfo {pages} {521} (\bibinfo {year} {2004})},\ \bibinfo {note}
  {[Erratum: Phys.Lett.B 632, 754--756 (2006)]},\ \Eprint
  {https://arxiv.org/abs/hep-th/0404092} {arXiv:hep-th/0404092} \BibitemShut
  {NoStop}%
\bibitem [{\citenamefont {Del~Duca}\ \emph {et~al.}(2010)\citenamefont
  {Del~Duca}, \citenamefont {Duhr},\ and\ \citenamefont
  {Smirnov}}]{DelDuca:2010zg}%
  \BibitemOpen
  \bibfield  {author} {\bibinfo {author} {\bibfnamefont {V.}~\bibnamefont
  {Del~Duca}}, \bibinfo {author} {\bibfnamefont {C.}~\bibnamefont {Duhr}},\
  and\ \bibinfo {author} {\bibfnamefont {V.~A.}\ \bibnamefont {Smirnov}},\
  }\bibfield  {title} {\bibinfo {title} {{The Two-Loop Hexagon Wilson Loop in N
  = 4 SYM}},\ }\href {https://doi.org/10.1007/JHEP05(2010)084} {\bibfield
  {journal} {\bibinfo  {journal} {JHEP}\ }\textbf {\bibinfo {volume} {05}},\
  \bibinfo {pages} {084}},\ \Eprint {https://arxiv.org/abs/1003.1702}
  {arXiv:1003.1702 [hep-th]} \BibitemShut {NoStop}%
\bibitem [{\citenamefont {Duhr}\ \emph {et~al.}(2012)\citenamefont {Duhr},
  \citenamefont {Gangl},\ and\ \citenamefont {Rhodes}}]{Duhr:2011zq}%
  \BibitemOpen
  \bibfield  {author} {\bibinfo {author} {\bibfnamefont {C.}~\bibnamefont
  {Duhr}}, \bibinfo {author} {\bibfnamefont {H.}~\bibnamefont {Gangl}},\ and\
  \bibinfo {author} {\bibfnamefont {J.~R.}\ \bibnamefont {Rhodes}},\ }\bibfield
   {title} {\bibinfo {title} {{From polygons and symbols to polylogarithmic
  functions}},\ }\href {https://doi.org/10.1007/JHEP10(2012)075} {\bibfield
  {journal} {\bibinfo  {journal} {JHEP}\ }\textbf {\bibinfo {volume} {10}},\
  \bibinfo {pages} {075}},\ \Eprint {https://arxiv.org/abs/1110.0458}
  {arXiv:1110.0458 [math-ph]} \BibitemShut {NoStop}%
\bibitem [{\citenamefont {Dixon}\ and\ \citenamefont
  {Liu}(2023)}]{Dixon:2023kop}%
  \BibitemOpen
  \bibfield  {author} {\bibinfo {author} {\bibfnamefont {L.~J.}\ \bibnamefont
  {Dixon}}\ and\ \bibinfo {author} {\bibfnamefont {Y.-T.}\ \bibnamefont
  {Liu}},\ }\bibfield  {title} {\bibinfo {title} {{An eight loop amplitude via
  antipodal duality}},\ }\href {https://doi.org/10.1007/JHEP09(2023)098}
  {\bibfield  {journal} {\bibinfo  {journal} {JHEP}\ }\textbf {\bibinfo
  {volume} {09}},\ \bibinfo {pages} {098}},\ \Eprint
  {https://arxiv.org/abs/2308.08199} {arXiv:2308.08199 [hep-th]} \BibitemShut
  {NoStop}%
\bibitem [{\citenamefont {Dixon}\ and\ \citenamefont
  {Duhr}(2025)}]{Dixon:2025zwj}%
  \BibitemOpen
  \bibfield  {author} {\bibinfo {author} {\bibfnamefont {L.~J.}\ \bibnamefont
  {Dixon}}\ and\ \bibinfo {author} {\bibfnamefont {C.}~\bibnamefont {Duhr}},\
  }\bibfield  {title} {\bibinfo {title} {{Antipodal self-duality of square
  fishnet graphs}},\ }\href {https://doi.org/10.1103/PhysRevD.111.L101901}
  {\bibfield  {journal} {\bibinfo  {journal} {Phys. Rev. D}\ }\textbf {\bibinfo
  {volume} {111}},\ \bibinfo {pages} {L101901} (\bibinfo {year} {2025})},\
  \Eprint {https://arxiv.org/abs/2502.00862} {arXiv:2502.00862 [hep-th]}
  \BibitemShut {NoStop}%
\bibitem [{\citenamefont {Cai}\ \emph {et~al.}(2025)\citenamefont {Cai},
  \citenamefont {Charton}, \citenamefont {Cranmer}, \citenamefont {Dixon},
  \citenamefont {Merz},\ and\ \citenamefont {Wilhelm}}]{Cai:2025atc}%
  \BibitemOpen
  \bibfield  {author} {\bibinfo {author} {\bibfnamefont {T.}~\bibnamefont
  {Cai}}, \bibinfo {author} {\bibfnamefont {F.}~\bibnamefont {Charton}},
  \bibinfo {author} {\bibfnamefont {K.}~\bibnamefont {Cranmer}}, \bibinfo
  {author} {\bibfnamefont {L.~J.}\ \bibnamefont {Dixon}}, \bibinfo {author}
  {\bibfnamefont {G.~W.}\ \bibnamefont {Merz}},\ and\ \bibinfo {author}
  {\bibfnamefont {M.}~\bibnamefont {Wilhelm}},\ }\bibfield  {title} {\bibinfo
  {title} {{Recurrent features of amplitudes in planar $ \mathcal{N} $ = 4
  super Yang-Mills theory}},\ }\href {https://doi.org/10.1007/JHEP04(2025)143}
  {\bibfield  {journal} {\bibinfo  {journal} {JHEP}\ }\textbf {\bibinfo
  {volume} {04}},\ \bibinfo {pages} {143}},\ \Eprint
  {https://arxiv.org/abs/2501.05743} {arXiv:2501.05743 [hep-th]} \BibitemShut
  {NoStop}%
\bibitem [{\citenamefont {Cai}\ \emph {et~al.}(2024)\citenamefont {Cai},
  \citenamefont {Merz}, \citenamefont {Charton}, \citenamefont {Nolte},
  \citenamefont {Wilhelm}, \citenamefont {Cranmer},\ and\ \citenamefont
  {Dixon}}]{Cai:2024znx}%
  \BibitemOpen
  \bibfield  {author} {\bibinfo {author} {\bibfnamefont {T.}~\bibnamefont
  {Cai}}, \bibinfo {author} {\bibfnamefont {G.~W.}\ \bibnamefont {Merz}},
  \bibinfo {author} {\bibfnamefont {F.}~\bibnamefont {Charton}}, \bibinfo
  {author} {\bibfnamefont {N.}~\bibnamefont {Nolte}}, \bibinfo {author}
  {\bibfnamefont {M.}~\bibnamefont {Wilhelm}}, \bibinfo {author} {\bibfnamefont
  {K.}~\bibnamefont {Cranmer}},\ and\ \bibinfo {author} {\bibfnamefont {L.~J.}\
  \bibnamefont {Dixon}},\ }\bibfield  {title} {\bibinfo {title} {{Transforming
  the bootstrap: using transformers to compute scattering amplitudes in planar
  $\mathcal{N} = 4$ super Yang{\textendash}Mills theory}},\ }\href
  {https://doi.org/10.1088/2632-2153/ad743e} {\bibfield  {journal} {\bibinfo
  {journal} {Mach. Learn. Sci. Tech.}\ }\textbf {\bibinfo {volume} {5}},\
  \bibinfo {pages} {035073} (\bibinfo {year} {2024})},\ \Eprint
  {https://arxiv.org/abs/2405.06107} {arXiv:2405.06107 [cs.LG]} \BibitemShut
  {NoStop}%
\bibitem [{\citenamefont {Dixon}\ \emph {et~al.}(2023)\citenamefont {Dixon},
  \citenamefont {G{\"u}rdo{\u{g}}an}, \citenamefont {Liu}, \citenamefont
  {McLeod},\ and\ \citenamefont {Wilhelm}}]{Dixon:2022xqh}%
  \BibitemOpen
  \bibfield  {author} {\bibinfo {author} {\bibfnamefont {L.~J.}\ \bibnamefont
  {Dixon}}, \bibinfo {author} {\bibfnamefont {{\"O}.}~\bibnamefont
  {G{\"u}rdo{\u{g}}an}}, \bibinfo {author} {\bibfnamefont {Y.-T.}\ \bibnamefont
  {Liu}}, \bibinfo {author} {\bibfnamefont {A.~J.}\ \bibnamefont {McLeod}},\
  and\ \bibinfo {author} {\bibfnamefont {M.}~\bibnamefont {Wilhelm}},\
  }\bibfield  {title} {\bibinfo {title} {{Antipodal Self-Duality for a
  Four-Particle Form Factor}},\ }\href
  {https://doi.org/10.1103/PhysRevLett.130.111601} {\bibfield  {journal}
  {\bibinfo  {journal} {Phys. Rev. Lett.}\ }\textbf {\bibinfo {volume} {130}},\
  \bibinfo {pages} {111601} (\bibinfo {year} {2023})},\ \Eprint
  {https://arxiv.org/abs/2212.02410} {arXiv:2212.02410 [hep-th]} \BibitemShut
  {NoStop}%
\bibitem [{\citenamefont {Dixon}\ \emph
  {et~al.}(2022{\natexlab{a}})\citenamefont {Dixon}, \citenamefont {Gurdogan},
  \citenamefont {McLeod},\ and\ \citenamefont {Wilhelm}}]{Dixon:2021tdw}%
  \BibitemOpen
  \bibfield  {author} {\bibinfo {author} {\bibfnamefont {L.~J.}\ \bibnamefont
  {Dixon}}, \bibinfo {author} {\bibfnamefont {O.}~\bibnamefont {Gurdogan}},
  \bibinfo {author} {\bibfnamefont {A.~J.}\ \bibnamefont {McLeod}},\ and\
  \bibinfo {author} {\bibfnamefont {M.}~\bibnamefont {Wilhelm}},\ }\bibfield
  {title} {\bibinfo {title} {{Folding Amplitudes into Form Factors: An
  Antipodal Duality}},\ }\href {https://doi.org/10.1103/PhysRevLett.128.111602}
  {\bibfield  {journal} {\bibinfo  {journal} {Phys. Rev. Lett.}\ }\textbf
  {\bibinfo {volume} {128}},\ \bibinfo {pages} {111602} (\bibinfo {year}
  {2022}{\natexlab{a}})},\ \Eprint {https://arxiv.org/abs/2112.06243}
  {arXiv:2112.06243 [hep-th]} \BibitemShut {NoStop}%
\bibitem [{\citenamefont {Dixon}\ \emph {et~al.}(2021)\citenamefont {Dixon},
  \citenamefont {McLeod},\ and\ \citenamefont {Wilhelm}}]{Dixon:2020bbt}%
  \BibitemOpen
  \bibfield  {author} {\bibinfo {author} {\bibfnamefont {L.~J.}\ \bibnamefont
  {Dixon}}, \bibinfo {author} {\bibfnamefont {A.~J.}\ \bibnamefont {McLeod}},\
  and\ \bibinfo {author} {\bibfnamefont {M.}~\bibnamefont {Wilhelm}},\
  }\bibfield  {title} {\bibinfo {title} {{A Three-Point Form Factor Through
  Five Loops}},\ }\href {https://doi.org/10.1007/JHEP04(2021)147} {\bibfield
  {journal} {\bibinfo  {journal} {JHEP}\ }\textbf {\bibinfo {volume} {04}},\
  \bibinfo {pages} {147}},\ \Eprint {https://arxiv.org/abs/2012.12286}
  {arXiv:2012.12286 [hep-th]} \BibitemShut {NoStop}%
\bibitem [{\citenamefont {Dixon}\ and\ \citenamefont
  {Liu}(2020)}]{Dixon:2020cnr}%
  \BibitemOpen
  \bibfield  {author} {\bibinfo {author} {\bibfnamefont {L.~J.}\ \bibnamefont
  {Dixon}}\ and\ \bibinfo {author} {\bibfnamefont {Y.-T.}\ \bibnamefont
  {Liu}},\ }\bibfield  {title} {\bibinfo {title} {{Lifting Heptagon Symbols to
  Functions}},\ }\href {https://doi.org/10.1007/JHEP10(2020)031} {\bibfield
  {journal} {\bibinfo  {journal} {JHEP}\ }\textbf {\bibinfo {volume} {10}},\
  \bibinfo {pages} {031}},\ \Eprint {https://arxiv.org/abs/2007.12966}
  {arXiv:2007.12966 [hep-th]} \BibitemShut {NoStop}%
\bibitem [{\citenamefont {Caron-Huot}\ \emph {et~al.}(2020)\citenamefont
  {Caron-Huot}, \citenamefont {Dixon}, \citenamefont {Drummond}, \citenamefont
  {Dulat}, \citenamefont {Foster}, \citenamefont {G{\"u}rdo{\u{g}}an},
  \citenamefont {von Hippel}, \citenamefont {McLeod},\ and\ \citenamefont
  {Papathanasiou}}]{Caron-Huot:2020bkp}%
  \BibitemOpen
  \bibfield  {author} {\bibinfo {author} {\bibfnamefont {S.}~\bibnamefont
  {Caron-Huot}}, \bibinfo {author} {\bibfnamefont {L.~J.}\ \bibnamefont
  {Dixon}}, \bibinfo {author} {\bibfnamefont {J.~M.}\ \bibnamefont {Drummond}},
  \bibinfo {author} {\bibfnamefont {F.}~\bibnamefont {Dulat}}, \bibinfo
  {author} {\bibfnamefont {J.}~\bibnamefont {Foster}}, \bibinfo {author}
  {\bibfnamefont {{\"O}.}~\bibnamefont {G{\"u}rdo{\u{g}}an}}, \bibinfo {author}
  {\bibfnamefont {M.}~\bibnamefont {von Hippel}}, \bibinfo {author}
  {\bibfnamefont {A.~J.}\ \bibnamefont {McLeod}},\ and\ \bibinfo {author}
  {\bibfnamefont {G.}~\bibnamefont {Papathanasiou}},\ }\bibfield  {title}
  {\bibinfo {title} {{The Steinmann Cluster Bootstrap for $N$ = 4 Super
  Yang-Mills Amplitudes}},\ }\href {https://doi.org/10.22323/1.376.0003}
  {\bibfield  {journal} {\bibinfo  {journal} {PoS}\ }\textbf {\bibinfo {volume}
  {CORFU2019}},\ \bibinfo {pages} {003} (\bibinfo {year} {2020})},\ \Eprint
  {https://arxiv.org/abs/2005.06735} {arXiv:2005.06735 [hep-th]} \BibitemShut
  {NoStop}%
\bibitem [{\citenamefont {Basso}\ \emph {et~al.}(2020)\citenamefont {Basso},
  \citenamefont {Dixon},\ and\ \citenamefont {Papathanasiou}}]{Basso:2020xts}%
  \BibitemOpen
  \bibfield  {author} {\bibinfo {author} {\bibfnamefont {B.}~\bibnamefont
  {Basso}}, \bibinfo {author} {\bibfnamefont {L.~J.}\ \bibnamefont {Dixon}},\
  and\ \bibinfo {author} {\bibfnamefont {G.}~\bibnamefont {Papathanasiou}},\
  }\bibfield  {title} {\bibinfo {title} {{Origin of the Six-Gluon Amplitude in
  Planar $N=4$ Supersymmetric Yang-Mills Theory}},\ }\href
  {https://doi.org/10.1103/PhysRevLett.124.161603} {\bibfield  {journal}
  {\bibinfo  {journal} {Phys. Rev. Lett.}\ }\textbf {\bibinfo {volume} {124}},\
  \bibinfo {pages} {161603} (\bibinfo {year} {2020})},\ \Eprint
  {https://arxiv.org/abs/2001.05460} {arXiv:2001.05460 [hep-th]} \BibitemShut
  {NoStop}%
\bibitem [{\citenamefont {Caron-Huot}\ \emph
  {et~al.}(2019{\natexlab{a}})\citenamefont {Caron-Huot}, \citenamefont
  {Dixon}, \citenamefont {Dulat}, \citenamefont {von Hippel}, \citenamefont
  {McLeod},\ and\ \citenamefont {Papathanasiou}}]{Caron-Huot:2019vjl}%
  \BibitemOpen
  \bibfield  {author} {\bibinfo {author} {\bibfnamefont {S.}~\bibnamefont
  {Caron-Huot}}, \bibinfo {author} {\bibfnamefont {L.~J.}\ \bibnamefont
  {Dixon}}, \bibinfo {author} {\bibfnamefont {F.}~\bibnamefont {Dulat}},
  \bibinfo {author} {\bibfnamefont {M.}~\bibnamefont {von Hippel}}, \bibinfo
  {author} {\bibfnamefont {A.~J.}\ \bibnamefont {McLeod}},\ and\ \bibinfo
  {author} {\bibfnamefont {G.}~\bibnamefont {Papathanasiou}},\ }\bibfield
  {title} {\bibinfo {title} {{Six-Gluon amplitudes in planar $ \mathcal{N} $ =
  4 super-Yang-Mills theory at six and seven loops}},\ }\href
  {https://doi.org/10.1007/JHEP08(2019)016} {\bibfield  {journal} {\bibinfo
  {journal} {JHEP}\ }\textbf {\bibinfo {volume} {08}},\ \bibinfo {pages}
  {016}},\ \Eprint {https://arxiv.org/abs/1903.10890} {arXiv:1903.10890
  [hep-th]} \BibitemShut {NoStop}%
\bibitem [{\citenamefont {Li}(2025)}]{Li:2024rkq}%
  \BibitemOpen
  \bibfield  {author} {\bibinfo {author} {\bibfnamefont {Z.}~\bibnamefont
  {Li}},\ }\bibfield  {title} {\bibinfo {title} {{Two-loop MHV form factors
  from the periodic Wilson loop}},\ }\href
  {https://doi.org/10.1007/JHEP05(2025)209} {\bibfield  {journal} {\bibinfo
  {journal} {JHEP}\ }\textbf {\bibinfo {volume} {05}},\ \bibinfo {pages}
  {209}},\ \Eprint {https://arxiv.org/abs/2412.17974} {arXiv:2412.17974
  [hep-th]} \BibitemShut {NoStop}%
\bibitem [{\citenamefont {Drummond}\ \emph {et~al.}(2015)\citenamefont
  {Drummond}, \citenamefont {Papathanasiou},\ and\ \citenamefont
  {Spradlin}}]{Drummond:2014ffa}%
  \BibitemOpen
  \bibfield  {author} {\bibinfo {author} {\bibfnamefont {J.~M.}\ \bibnamefont
  {Drummond}}, \bibinfo {author} {\bibfnamefont {G.}~\bibnamefont
  {Papathanasiou}},\ and\ \bibinfo {author} {\bibfnamefont {M.}~\bibnamefont
  {Spradlin}},\ }\bibfield  {title} {\bibinfo {title} {{A Symbol of Uniqueness:
  The Cluster Bootstrap for the 3-Loop MHV Heptagon}},\ }\href
  {https://doi.org/10.1007/JHEP03(2015)072} {\bibfield  {journal} {\bibinfo
  {journal} {JHEP}\ }\textbf {\bibinfo {volume} {03}},\ \bibinfo {pages}
  {072}},\ \Eprint {https://arxiv.org/abs/1412.3763} {arXiv:1412.3763 [hep-th]}
  \BibitemShut {NoStop}%
\bibitem [{\citenamefont {Yang}(2020)}]{Yang:2019vag}%
  \BibitemOpen
  \bibfield  {author} {\bibinfo {author} {\bibfnamefont {G.}~\bibnamefont
  {Yang}},\ }\bibfield  {title} {\bibinfo {title} {{On-shell methods for form
  factors in $\mathcal{N}=4$ SYM and their applications}},\ }\href
  {https://doi.org/10.1007/s11433-019-1507-0} {\bibfield  {journal} {\bibinfo
  {journal} {Sci. China Phys. Mech. Astron.}\ }\textbf {\bibinfo {volume}
  {63}},\ \bibinfo {pages} {270001} (\bibinfo {year} {2020})},\ \Eprint
  {https://arxiv.org/abs/1912.11454} {arXiv:1912.11454 [hep-th]} \BibitemShut
  {NoStop}%
\bibitem [{\citenamefont {Dixon}\ \emph
  {et~al.}(2022{\natexlab{b}})\citenamefont {Dixon}, \citenamefont {Gurdogan},
  \citenamefont {McLeod},\ and\ \citenamefont {Wilhelm}}]{Dixon:2022rse}%
  \BibitemOpen
  \bibfield  {author} {\bibinfo {author} {\bibfnamefont {L.~J.}\ \bibnamefont
  {Dixon}}, \bibinfo {author} {\bibfnamefont {O.}~\bibnamefont {Gurdogan}},
  \bibinfo {author} {\bibfnamefont {A.~J.}\ \bibnamefont {McLeod}},\ and\
  \bibinfo {author} {\bibfnamefont {M.}~\bibnamefont {Wilhelm}},\ }\bibfield
  {title} {\bibinfo {title} {{Bootstrapping a stress-tensor form factor through
  eight loops}},\ }\href {https://doi.org/10.1007/JHEP07(2022)153} {\bibfield
  {journal} {\bibinfo  {journal} {JHEP}\ }\textbf {\bibinfo {volume} {07}},\
  \bibinfo {pages} {153}},\ \Eprint {https://arxiv.org/abs/2204.11901}
  {arXiv:2204.11901 [hep-th]} \BibitemShut {NoStop}%
\bibitem [{\citenamefont {Gehrmann}\ \emph {et~al.}(2024)\citenamefont
  {Gehrmann}, \citenamefont {Henn}, \citenamefont {Jakub\v{c}\'\i{}k},
  \citenamefont {Lim}, \citenamefont {Mella}, \citenamefont {Syrrakos},
  \citenamefont {Tancredi},\ and\ \citenamefont
  {Torres~Bobadilla}}]{Gehrmann:2024tds}%
  \BibitemOpen
  \bibfield  {author} {\bibinfo {author} {\bibfnamefont {T.}~\bibnamefont
  {Gehrmann}}, \bibinfo {author} {\bibfnamefont {J.}~\bibnamefont {Henn}},
  \bibinfo {author} {\bibfnamefont {P.}~\bibnamefont {Jakub\v{c}\'\i{}k}},
  \bibinfo {author} {\bibfnamefont {J.}~\bibnamefont {Lim}}, \bibinfo {author}
  {\bibfnamefont {C.~C.}\ \bibnamefont {Mella}}, \bibinfo {author}
  {\bibfnamefont {N.}~\bibnamefont {Syrrakos}}, \bibinfo {author}
  {\bibfnamefont {L.}~\bibnamefont {Tancredi}},\ and\ \bibinfo {author}
  {\bibfnamefont {W.~J.}\ \bibnamefont {Torres~Bobadilla}},\ }\bibfield
  {title} {\bibinfo {title} {{Graded transcendental functions: an application
  to four-point amplitudes with one off-shell leg}},\ }\href
  {https://doi.org/10.1007/JHEP12(2024)215} {\bibfield  {journal} {\bibinfo
  {journal} {JHEP}\ }\textbf {\bibinfo {volume} {12}},\ \bibinfo {pages}
  {215}},\ \Eprint {https://arxiv.org/abs/2410.19088} {arXiv:2410.19088
  [hep-th]} \BibitemShut {NoStop}%
\bibitem [{\citenamefont {Dixon}\ and\ \citenamefont
  {Xin}(2025)}]{Dixon:2024yvq}%
  \BibitemOpen
  \bibfield  {author} {\bibinfo {author} {\bibfnamefont {L.~J.}\ \bibnamefont
  {Dixon}}\ and\ \bibinfo {author} {\bibfnamefont {S.}~\bibnamefont {Xin}},\
  }\bibfield  {title} {\bibinfo {title} {{A two-loop four-point form factor at
  function level}},\ }\href {https://doi.org/10.1007/JHEP01(2025)012}
  {\bibfield  {journal} {\bibinfo  {journal} {JHEP}\ }\textbf {\bibinfo
  {volume} {01}},\ \bibinfo {pages} {012}},\ \Eprint
  {https://arxiv.org/abs/2411.01571} {arXiv:2411.01571 [hep-th]} \BibitemShut
  {NoStop}%
\bibitem [{\citenamefont {Penante}\ \emph {et~al.}(2014)\citenamefont
  {Penante}, \citenamefont {Spence}, \citenamefont {Travaglini},\ and\
  \citenamefont {Wen}}]{Penante:2014sza}%
  \BibitemOpen
  \bibfield  {author} {\bibinfo {author} {\bibfnamefont {B.}~\bibnamefont
  {Penante}}, \bibinfo {author} {\bibfnamefont {B.}~\bibnamefont {Spence}},
  \bibinfo {author} {\bibfnamefont {G.}~\bibnamefont {Travaglini}},\ and\
  \bibinfo {author} {\bibfnamefont {C.}~\bibnamefont {Wen}},\ }\bibfield
  {title} {\bibinfo {title} {{On super form factors of half-BPS operators in
  N=4 super Yang-Mills}},\ }\href {https://doi.org/10.1007/JHEP04(2014)083}
  {\bibfield  {journal} {\bibinfo  {journal} {JHEP}\ }\textbf {\bibinfo
  {volume} {04}},\ \bibinfo {pages} {083}},\ \Eprint
  {https://arxiv.org/abs/1402.1300} {arXiv:1402.1300 [hep-th]} \BibitemShut
  {NoStop}%
\bibitem [{\citenamefont {Henn}\ \emph {et~al.}(2025)\citenamefont {Henn},
  \citenamefont {Lim},\ and\ \citenamefont {Torres~Bobadilla}}]{Henn:2024pki}%
  \BibitemOpen
  \bibfield  {author} {\bibinfo {author} {\bibfnamefont {J.~M.}\ \bibnamefont
  {Henn}}, \bibinfo {author} {\bibfnamefont {J.}~\bibnamefont {Lim}},\ and\
  \bibinfo {author} {\bibfnamefont {W.~J.}\ \bibnamefont {Torres~Bobadilla}},\
  }\bibfield  {title} {\bibinfo {title} {{Analytic evaluation of the three-loop
  three-point form factor of tr {\ensuremath{\phi}}$^{3}$ in $ \mathcal{N} $ =
  4 sYM}},\ }\href {https://doi.org/10.1007/JHEP02(2025)085} {\bibfield
  {journal} {\bibinfo  {journal} {JHEP}\ }\textbf {\bibinfo {volume} {02}},\
  \bibinfo {pages} {085}},\ \Eprint {https://arxiv.org/abs/2410.22465}
  {arXiv:2410.22465 [hep-th]} \BibitemShut {NoStop}%
\bibitem [{\citenamefont {Basso}\ \emph {et~al.}(2025)\citenamefont {Basso},
  \citenamefont {Dixon},\ and\ \citenamefont {Tumanov}}]{Basso:2024hlx}%
  \BibitemOpen
  \bibfield  {author} {\bibinfo {author} {\bibfnamefont {B.}~\bibnamefont
  {Basso}}, \bibinfo {author} {\bibfnamefont {L.~J.}\ \bibnamefont {Dixon}},\
  and\ \bibinfo {author} {\bibfnamefont {A.~G.}\ \bibnamefont {Tumanov}},\
  }\bibfield  {title} {\bibinfo {title} {{The three-point form factor of Tr
  {\ensuremath{\phi}}$^{3}$ to six loops}},\ }\href
  {https://doi.org/10.1007/JHEP02(2025)034} {\bibfield  {journal} {\bibinfo
  {journal} {JHEP}\ }\textbf {\bibinfo {volume} {02}},\ \bibinfo {pages}
  {034}},\ \Eprint {https://arxiv.org/abs/2410.22402} {arXiv:2410.22402
  [hep-th]} \BibitemShut {NoStop}%
\bibitem [{\citenamefont {Anastasiou}\ \emph {et~al.}(2003)\citenamefont
  {Anastasiou}, \citenamefont {Bern}, \citenamefont {Dixon},\ and\
  \citenamefont {Kosower}}]{Anastasiou:2003kj}%
  \BibitemOpen
  \bibfield  {author} {\bibinfo {author} {\bibfnamefont {C.}~\bibnamefont
  {Anastasiou}}, \bibinfo {author} {\bibfnamefont {Z.}~\bibnamefont {Bern}},
  \bibinfo {author} {\bibfnamefont {L.~J.}\ \bibnamefont {Dixon}},\ and\
  \bibinfo {author} {\bibfnamefont {D.~A.}\ \bibnamefont {Kosower}},\
  }\bibfield  {title} {\bibinfo {title} {{Planar amplitudes in maximally
  supersymmetric Yang-Mills theory}},\ }\href
  {https://doi.org/10.1103/PhysRevLett.91.251602} {\bibfield  {journal}
  {\bibinfo  {journal} {Phys. Rev. Lett.}\ }\textbf {\bibinfo {volume} {91}},\
  \bibinfo {pages} {251602} (\bibinfo {year} {2003})},\ \Eprint
  {https://arxiv.org/abs/hep-th/0309040} {arXiv:hep-th/0309040} \BibitemShut
  {NoStop}%
\bibitem [{\citenamefont {Bern}\ \emph {et~al.}(2005)\citenamefont {Bern},
  \citenamefont {Dixon},\ and\ \citenamefont {Smirnov}}]{Bern:2005iz}%
  \BibitemOpen
  \bibfield  {author} {\bibinfo {author} {\bibfnamefont {Z.}~\bibnamefont
  {Bern}}, \bibinfo {author} {\bibfnamefont {L.~J.}\ \bibnamefont {Dixon}},\
  and\ \bibinfo {author} {\bibfnamefont {V.~A.}\ \bibnamefont {Smirnov}},\
  }\bibfield  {title} {\bibinfo {title} {{Iteration of planar amplitudes in
  maximally supersymmetric Yang-Mills theory at three loops and beyond}},\
  }\href {https://doi.org/10.1103/PhysRevD.72.085001} {\bibfield  {journal}
  {\bibinfo  {journal} {Phys. Rev. D}\ }\textbf {\bibinfo {volume} {72}},\
  \bibinfo {pages} {085001} (\bibinfo {year} {2005})},\ \Eprint
  {https://arxiv.org/abs/hep-th/0505205} {arXiv:hep-th/0505205} \BibitemShut
  {NoStop}%
\bibitem [{\citenamefont {Henn}\ and\ \citenamefont
  {Mistlberger}(2016)}]{Henn:2016jdu}%
  \BibitemOpen
  \bibfield  {author} {\bibinfo {author} {\bibfnamefont {J.~M.}\ \bibnamefont
  {Henn}}\ and\ \bibinfo {author} {\bibfnamefont {B.}~\bibnamefont
  {Mistlberger}},\ }\bibfield  {title} {\bibinfo {title} {{Four-Gluon
  Scattering at Three Loops, Infrared Structure, and the Regge Limit}},\ }\href
  {https://doi.org/10.1103/PhysRevLett.117.171601} {\bibfield  {journal}
  {\bibinfo  {journal} {Phys. Rev. Lett.}\ }\textbf {\bibinfo {volume} {117}},\
  \bibinfo {pages} {171601} (\bibinfo {year} {2016})},\ \Eprint
  {https://arxiv.org/abs/1608.00850} {arXiv:1608.00850 [hep-th]} \BibitemShut
  {NoStop}%
\bibitem [{\citenamefont {Chicherin}\ \emph {et~al.}(2019)\citenamefont
  {Chicherin}, \citenamefont {Gehrmann}, \citenamefont {Henn}, \citenamefont
  {Wasser}, \citenamefont {Zhang},\ and\ \citenamefont
  {Zoia}}]{Chicherin:2018old}%
  \BibitemOpen
  \bibfield  {author} {\bibinfo {author} {\bibfnamefont {D.}~\bibnamefont
  {Chicherin}}, \bibinfo {author} {\bibfnamefont {T.}~\bibnamefont {Gehrmann}},
  \bibinfo {author} {\bibfnamefont {J.~M.}\ \bibnamefont {Henn}}, \bibinfo
  {author} {\bibfnamefont {P.}~\bibnamefont {Wasser}}, \bibinfo {author}
  {\bibfnamefont {Y.}~\bibnamefont {Zhang}},\ and\ \bibinfo {author}
  {\bibfnamefont {S.}~\bibnamefont {Zoia}},\ }\bibfield  {title} {\bibinfo
  {title} {{All Master Integrals for Three-Jet Production at
  Next-to-Next-to-Leading Order}},\ }\href
  {https://doi.org/10.1103/PhysRevLett.123.041603} {\bibfield  {journal}
  {\bibinfo  {journal} {Phys. Rev. Lett.}\ }\textbf {\bibinfo {volume} {123}},\
  \bibinfo {pages} {041603} (\bibinfo {year} {2019})},\ \Eprint
  {https://arxiv.org/abs/1812.11160} {arXiv:1812.11160 [hep-ph]} \BibitemShut
  {NoStop}%
\bibitem [{\citenamefont {Abreu}\ \emph {et~al.}(2019)\citenamefont {Abreu},
  \citenamefont {Dixon}, \citenamefont {Herrmann}, \citenamefont {Page},\ and\
  \citenamefont {Zeng}}]{Abreu:2018aqd}%
  \BibitemOpen
  \bibfield  {author} {\bibinfo {author} {\bibfnamefont {S.}~\bibnamefont
  {Abreu}}, \bibinfo {author} {\bibfnamefont {L.~J.}\ \bibnamefont {Dixon}},
  \bibinfo {author} {\bibfnamefont {E.}~\bibnamefont {Herrmann}}, \bibinfo
  {author} {\bibfnamefont {B.}~\bibnamefont {Page}},\ and\ \bibinfo {author}
  {\bibfnamefont {M.}~\bibnamefont {Zeng}},\ }\bibfield  {title} {\bibinfo
  {title} {{The two-loop five-point amplitude in $\mathcal{N} =4$
  super-Yang-Mills theory}},\ }\href
  {https://doi.org/10.1103/PhysRevLett.122.121603} {\bibfield  {journal}
  {\bibinfo  {journal} {Phys. Rev. Lett.}\ }\textbf {\bibinfo {volume} {122}},\
  \bibinfo {pages} {121603} (\bibinfo {year} {2019})},\ \Eprint
  {https://arxiv.org/abs/1812.08941} {arXiv:1812.08941 [hep-th]} \BibitemShut
  {NoStop}%
\bibitem [{\citenamefont {Brandhuber}\ \emph {et~al.}(2012)\citenamefont
  {Brandhuber}, \citenamefont {Travaglini},\ and\ \citenamefont
  {Yang}}]{Brandhuber:2012vm}%
  \BibitemOpen
  \bibfield  {author} {\bibinfo {author} {\bibfnamefont {A.}~\bibnamefont
  {Brandhuber}}, \bibinfo {author} {\bibfnamefont {G.}~\bibnamefont
  {Travaglini}},\ and\ \bibinfo {author} {\bibfnamefont {G.}~\bibnamefont
  {Yang}},\ }\bibfield  {title} {\bibinfo {title} {{Analytic two-loop form
  factors in N=4 SYM}},\ }\href {https://doi.org/10.1007/JHEP05(2012)082}
  {\bibfield  {journal} {\bibinfo  {journal} {JHEP}\ }\textbf {\bibinfo
  {volume} {05}},\ \bibinfo {pages} {082}},\ \Eprint
  {https://arxiv.org/abs/1201.4170} {arXiv:1201.4170 [hep-th]} \BibitemShut
  {NoStop}%
\bibitem [{\citenamefont {Chen}\ \emph {et~al.}(2025)\citenamefont {Chen},
  \citenamefont {Guan},\ and\ \citenamefont {Mistlberger}}]{Chen:2025utl}%
  \BibitemOpen
  \bibfield  {author} {\bibinfo {author} {\bibfnamefont {X.}~\bibnamefont
  {Chen}}, \bibinfo {author} {\bibfnamefont {X.}~\bibnamefont {Guan}},\ and\
  \bibinfo {author} {\bibfnamefont {B.}~\bibnamefont {Mistlberger}},\
  }\bibfield  {title} {\bibinfo {title} {{Three-Loop QCD corrections to the
  production of a Higgs boson and a Jet}},\ }\href@noop {} {\  (\bibinfo {year}
  {2025})},\ \Eprint {https://arxiv.org/abs/2504.06490} {arXiv:2504.06490
  [hep-ph]} \BibitemShut {NoStop}%
\bibitem [{\citenamefont {Lin}\ \emph {et~al.}(2021)\citenamefont {Lin},
  \citenamefont {Yang},\ and\ \citenamefont {Zhang}}]{Lin:2021kht}%
  \BibitemOpen
  \bibfield  {author} {\bibinfo {author} {\bibfnamefont {G.}~\bibnamefont
  {Lin}}, \bibinfo {author} {\bibfnamefont {G.}~\bibnamefont {Yang}},\ and\
  \bibinfo {author} {\bibfnamefont {S.}~\bibnamefont {Zhang}},\ }\bibfield
  {title} {\bibinfo {title} {{Three-Loop Color-Kinematics Duality: A
  24-Dimensional Solution Space Induced by New Generalized Gauge
  Transformations}},\ }\href {https://doi.org/10.1103/PhysRevLett.127.171602}
  {\bibfield  {journal} {\bibinfo  {journal} {Phys. Rev. Lett.}\ }\textbf
  {\bibinfo {volume} {127}},\ \bibinfo {pages} {171602} (\bibinfo {year}
  {2021})},\ \Eprint {https://arxiv.org/abs/2106.05280} {arXiv:2106.05280
  [hep-th]} \BibitemShut {NoStop}%
\bibitem [{\citenamefont {Lin}\ \emph {et~al.}(2022)\citenamefont {Lin},
  \citenamefont {Yang},\ and\ \citenamefont {Zhang}}]{Lin:2021qol}%
  \BibitemOpen
  \bibfield  {author} {\bibinfo {author} {\bibfnamefont {G.}~\bibnamefont
  {Lin}}, \bibinfo {author} {\bibfnamefont {G.}~\bibnamefont {Yang}},\ and\
  \bibinfo {author} {\bibfnamefont {S.}~\bibnamefont {Zhang}},\ }\bibfield
  {title} {\bibinfo {title} {{Full-color three-loop three-point form factors in
  N = 4 SYM}},\ }\href {https://doi.org/10.1007/JHEP03(2022)061} {\bibfield
  {journal} {\bibinfo  {journal} {JHEP}\ }\textbf {\bibinfo {volume} {03}},\
  \bibinfo {pages} {061}},\ \Eprint {https://arxiv.org/abs/2111.03021}
  {arXiv:2111.03021 [hep-th]} \BibitemShut {NoStop}%
\bibitem [{\citenamefont {Guan}\ \emph {et~al.}(2024)\citenamefont {Guan},
  \citenamefont {Lin}, \citenamefont {Liu}, \citenamefont {Ma},\ and\
  \citenamefont {Yang}}]{Guan:2023gsz}%
  \BibitemOpen
  \bibfield  {author} {\bibinfo {author} {\bibfnamefont {X.}~\bibnamefont
  {Guan}}, \bibinfo {author} {\bibfnamefont {G.}~\bibnamefont {Lin}}, \bibinfo
  {author} {\bibfnamefont {X.}~\bibnamefont {Liu}}, \bibinfo {author}
  {\bibfnamefont {Y.-Q.}\ \bibnamefont {Ma}},\ and\ \bibinfo {author}
  {\bibfnamefont {G.}~\bibnamefont {Yang}},\ }\bibfield  {title} {\bibinfo
  {title} {{A high-precision result for a full-color three-loop three-point
  form factor in $ \mathcal{N} $ = 4 SYM}},\ }\href
  {https://doi.org/10.1007/JHEP02(2024)201} {\bibfield  {journal} {\bibinfo
  {journal} {JHEP}\ }\textbf {\bibinfo {volume} {02}},\ \bibinfo {pages}
  {201}},\ \Eprint {https://arxiv.org/abs/2309.04395} {arXiv:2309.04395
  [hep-th]} \BibitemShut {NoStop}%
\bibitem [{\citenamefont {Guan}\ \emph
  {et~al.}(2025{\natexlab{a}})\citenamefont {Guan}, \citenamefont {Liu},
  \citenamefont {Ma},\ and\ \citenamefont {Wu}}]{Guan:2024byi}%
  \BibitemOpen
  \bibfield  {author} {\bibinfo {author} {\bibfnamefont {X.}~\bibnamefont
  {Guan}}, \bibinfo {author} {\bibfnamefont {X.}~\bibnamefont {Liu}}, \bibinfo
  {author} {\bibfnamefont {Y.-Q.}\ \bibnamefont {Ma}},\ and\ \bibinfo {author}
  {\bibfnamefont {W.-H.}\ \bibnamefont {Wu}},\ }\bibfield  {title} {\bibinfo
  {title} {{Blade: A package for block-triangular form improved Feynman
  integrals decomposition}},\ }\href
  {https://doi.org/10.1016/j.cpc.2025.109538} {\bibfield  {journal} {\bibinfo
  {journal} {Comput. Phys. Commun.}\ }\textbf {\bibinfo {volume} {310}},\
  \bibinfo {pages} {109538} (\bibinfo {year} {2025}{\natexlab{a}})},\ \Eprint
  {https://arxiv.org/abs/2405.14621} {arXiv:2405.14621 [hep-ph]} \BibitemShut
  {NoStop}%
\bibitem [{\citenamefont {Tkachov}(1981)}]{Tkachov1981}%
  \BibitemOpen
  \bibfield  {author} {\bibinfo {author} {\bibfnamefont {F.}~\bibnamefont
  {Tkachov}},\ }\bibfield  {title} {\bibinfo {title} {{A theorem on analytical
  calculability of 4-loop renormalization group functions}},\ }\href
  {https://doi.org/10.1016/0370-2693(81)90288-4} {\bibfield  {journal}
  {\bibinfo  {journal} {Physics Letters B}\ }\textbf {\bibinfo {volume}
  {100}},\ \bibinfo {pages} {65} (\bibinfo {year} {1981})}\BibitemShut
  {NoStop}%
\bibitem [{\citenamefont {Chetyrkin}\ and\ \citenamefont
  {Tkachov}(1981)}]{Chetyrkin1981}%
  \BibitemOpen
  \bibfield  {author} {\bibinfo {author} {\bibfnamefont {K.}~\bibnamefont
  {Chetyrkin}}\ and\ \bibinfo {author} {\bibfnamefont {F.}~\bibnamefont
  {Tkachov}},\ }\bibfield  {title} {\bibinfo {title} {{Integration by parts:
  The algorithm to calculate $\beta$-functions in 4 loops}},\ }\href
  {https://doi.org/10.1016/0550-3213(81)90199-1} {\bibfield  {journal}
  {\bibinfo  {journal} {Nuclear Physics B}\ }\textbf {\bibinfo {volume}
  {192}},\ \bibinfo {pages} {159} (\bibinfo {year} {1981})}\BibitemShut
  {NoStop}%
\bibitem [{\citenamefont {Laporta}(2000)}]{Laporta:2001dd}%
  \BibitemOpen
  \bibfield  {author} {\bibinfo {author} {\bibfnamefont {S.}~\bibnamefont
  {Laporta}},\ }\bibfield  {title} {\bibinfo {title} {{High precision
  calculation of multiloop Feynman integrals by difference equations}},\ }\href
  {https://doi.org/10.1016/S0217-751X(00)00215-7, 10.1142/S0217751X00002157}
  {\bibfield  {journal} {\bibinfo  {journal} {Int. J. Mod. Phys.}\ }\textbf
  {\bibinfo {volume} {A15}},\ \bibinfo {pages} {5087} (\bibinfo {year}
  {2000})},\ \Eprint {https://arxiv.org/abs/hep-ph/0102033}
  {arXiv:hep-ph/0102033 [hep-ph]} \BibitemShut {NoStop}%
\bibitem [{\citenamefont {Liu}\ and\ \citenamefont {Ma}(2019)}]{Liu:2018dmc}%
  \BibitemOpen
  \bibfield  {author} {\bibinfo {author} {\bibfnamefont {X.}~\bibnamefont
  {Liu}}\ and\ \bibinfo {author} {\bibfnamefont {Y.-Q.}\ \bibnamefont {Ma}},\
  }\bibfield  {title} {\bibinfo {title} {{Determining arbitrary Feynman
  integrals by vacuum integrals}},\ }\href
  {https://doi.org/10.1103/PhysRevD.99.071501} {\bibfield  {journal} {\bibinfo
  {journal} {Phys. Rev. D}\ }\textbf {\bibinfo {volume} {99}},\ \bibinfo
  {pages} {071501} (\bibinfo {year} {2019})},\ \Eprint
  {https://arxiv.org/abs/1801.10523} {arXiv:1801.10523 [hep-ph]} \BibitemShut
  {NoStop}%
\bibitem [{\citenamefont {Guan}\ \emph {et~al.}(2020)\citenamefont {Guan},
  \citenamefont {Liu},\ and\ \citenamefont {Ma}}]{Guan:2019bcx}%
  \BibitemOpen
  \bibfield  {author} {\bibinfo {author} {\bibfnamefont {X.}~\bibnamefont
  {Guan}}, \bibinfo {author} {\bibfnamefont {X.}~\bibnamefont {Liu}},\ and\
  \bibinfo {author} {\bibfnamefont {Y.-Q.}\ \bibnamefont {Ma}},\ }\bibfield
  {title} {\bibinfo {title} {{Complete reduction of integrals in two-loop
  five-light-parton scattering amplitudes}},\ }\href
  {https://doi.org/10.1088/1674-1137/44/9/093106} {\bibfield  {journal}
  {\bibinfo  {journal} {Chin. Phys. C}\ }\textbf {\bibinfo {volume} {44}},\
  \bibinfo {pages} {093106} (\bibinfo {year} {2020})},\ \Eprint
  {https://arxiv.org/abs/1912.09294} {arXiv:1912.09294 [hep-ph]} \BibitemShut
  {NoStop}%
\bibitem [{\citenamefont {Di~Vita}\ \emph {et~al.}(2014)\citenamefont
  {Di~Vita}, \citenamefont {Mastrolia}, \citenamefont {Schubert},\ and\
  \citenamefont {Yundin}}]{DiVita:2014pza}%
  \BibitemOpen
  \bibfield  {author} {\bibinfo {author} {\bibfnamefont {S.}~\bibnamefont
  {Di~Vita}}, \bibinfo {author} {\bibfnamefont {P.}~\bibnamefont {Mastrolia}},
  \bibinfo {author} {\bibfnamefont {U.}~\bibnamefont {Schubert}},\ and\
  \bibinfo {author} {\bibfnamefont {V.}~\bibnamefont {Yundin}},\ }\bibfield
  {title} {\bibinfo {title} {{Three-loop master integrals for ladder-box
  diagrams with one massive leg}},\ }\href
  {https://doi.org/10.1007/JHEP09(2014)148} {\bibfield  {journal} {\bibinfo
  {journal} {JHEP}\ }\textbf {\bibinfo {volume} {09}},\ \bibinfo {pages}
  {148}},\ \Eprint {https://arxiv.org/abs/1408.3107} {arXiv:1408.3107 [hep-ph]}
  \BibitemShut {NoStop}%
\bibitem [{\citenamefont {Canko}\ and\ \citenamefont
  {Syrrakos}(2022)}]{Canko:2021xmn}%
  \BibitemOpen
  \bibfield  {author} {\bibinfo {author} {\bibfnamefont {D.~D.}\ \bibnamefont
  {Canko}}\ and\ \bibinfo {author} {\bibfnamefont {N.}~\bibnamefont
  {Syrrakos}},\ }\bibfield  {title} {\bibinfo {title} {{Planar three-loop
  master integrals for 2 \textrightarrow{} 2 processes with one external
  massive particle}},\ }\href {https://doi.org/10.1007/JHEP04(2022)134}
  {\bibfield  {journal} {\bibinfo  {journal} {JHEP}\ }\textbf {\bibinfo
  {volume} {04}},\ \bibinfo {pages} {134}},\ \Eprint
  {https://arxiv.org/abs/2112.14275} {arXiv:2112.14275 [hep-ph]} \BibitemShut
  {NoStop}%
\bibitem [{\citenamefont {Henn}\ \emph {et~al.}(2023)\citenamefont {Henn},
  \citenamefont {Lim},\ and\ \citenamefont {Torres~Bobadilla}}]{Henn:2023vbd}%
  \BibitemOpen
  \bibfield  {author} {\bibinfo {author} {\bibfnamefont {J.~M.}\ \bibnamefont
  {Henn}}, \bibinfo {author} {\bibfnamefont {J.}~\bibnamefont {Lim}},\ and\
  \bibinfo {author} {\bibfnamefont {W.~J.}\ \bibnamefont {Torres~Bobadilla}},\
  }\bibfield  {title} {\bibinfo {title} {{First look at the evaluation of
  three-loop non-planar Feynman diagrams for Higgs plus jet production}},\
  }\href {https://doi.org/10.1007/JHEP05(2023)026} {\bibfield  {journal}
  {\bibinfo  {journal} {JHEP}\ }\textbf {\bibinfo {volume} {05}},\ \bibinfo
  {pages} {026}},\ \Eprint {https://arxiv.org/abs/2302.12776} {arXiv:2302.12776
  [hep-th]} \BibitemShut {NoStop}%
\bibitem [{\citenamefont {Syrrakos}\ and\ \citenamefont
  {Canko}(2024)}]{Syrrakos:2023mor}%
  \BibitemOpen
  \bibfield  {author} {\bibinfo {author} {\bibfnamefont {N.}~\bibnamefont
  {Syrrakos}}\ and\ \bibinfo {author} {\bibfnamefont {D.~D.}\ \bibnamefont
  {Canko}},\ }\bibfield  {title} {\bibinfo {title} {{Three-loop master
  integrals for H+jet production at N3LO: Towards the non-planar topologies}},\
  }\href {https://doi.org/10.22323/1.432.0044} {\bibfield  {journal} {\bibinfo
  {journal} {PoS}\ }\textbf {\bibinfo {volume} {RADCOR2023}},\ \bibinfo {pages}
  {044} (\bibinfo {year} {2024})},\ \Eprint {https://arxiv.org/abs/2307.08432}
  {arXiv:2307.08432 [hep-ph]} \BibitemShut {NoStop}%
\bibitem [{\citenamefont {Kotikov}(1991{\natexlab{a}})}]{Kotikov:1990kg}%
  \BibitemOpen
  \bibfield  {author} {\bibinfo {author} {\bibfnamefont {A.}~\bibnamefont
  {Kotikov}},\ }\bibfield  {title} {\bibinfo {title} {{Differential equations
  method: New technique for massive Feynman diagrams calculation}},\ }\href
  {https://doi.org/10.1016/0370-2693(91)90413-K} {\bibfield  {journal}
  {\bibinfo  {journal} {Phys. Lett. B}\ }\textbf {\bibinfo {volume} {254}},\
  \bibinfo {pages} {158} (\bibinfo {year} {1991}{\natexlab{a}})}\BibitemShut
  {NoStop}%
\bibitem [{\citenamefont {Kotikov}(1991{\natexlab{b}})}]{Kotikov:1991hm}%
  \BibitemOpen
  \bibfield  {author} {\bibinfo {author} {\bibfnamefont {A.}~\bibnamefont
  {Kotikov}},\ }\bibfield  {title} {\bibinfo {title} {{Differential equations
  method: The Calculation of vertex type Feynman diagrams}},\ }\href
  {https://doi.org/10.1016/0370-2693(91)90834-D} {\bibfield  {journal}
  {\bibinfo  {journal} {Phys. Lett. B}\ }\textbf {\bibinfo {volume} {259}},\
  \bibinfo {pages} {314} (\bibinfo {year} {1991}{\natexlab{b}})}\BibitemShut
  {NoStop}%
\bibitem [{\citenamefont {Kotikov}(1991{\natexlab{c}})}]{Kotikov:1991pm}%
  \BibitemOpen
  \bibfield  {author} {\bibinfo {author} {\bibfnamefont {A.}~\bibnamefont
  {Kotikov}},\ }\bibfield  {title} {\bibinfo {title} {{Differential equation
  method: The Calculation of N point Feynman diagrams}},\ }\href
  {https://doi.org/10.1016/0370-2693(91)90536-Y} {\bibfield  {journal}
  {\bibinfo  {journal} {Phys. Lett. B}\ }\textbf {\bibinfo {volume} {267}},\
  \bibinfo {pages} {123} (\bibinfo {year} {1991}{\natexlab{c}})},\ \bibinfo
  {note} {[Erratum: Phys.Lett.B 295, 409--409 (1992)]}\BibitemShut {NoStop}%
\bibitem [{\citenamefont {Gehrmann}\ and\ \citenamefont
  {Remiddi}(2000)}]{Gehrmann:1999as}%
  \BibitemOpen
  \bibfield  {author} {\bibinfo {author} {\bibfnamefont {T.}~\bibnamefont
  {Gehrmann}}\ and\ \bibinfo {author} {\bibfnamefont {E.}~\bibnamefont
  {Remiddi}},\ }\bibfield  {title} {\bibinfo {title} {{Differential equations
  for two loop four point functions}},\ }\href
  {https://doi.org/10.1016/S0550-3213(00)00223-6} {\bibfield  {journal}
  {\bibinfo  {journal} {Nucl. Phys.}\ }\textbf {\bibinfo {volume} {B580}},\
  \bibinfo {pages} {485} (\bibinfo {year} {2000})},\ \Eprint
  {https://arxiv.org/abs/hep-ph/9912329} {arXiv:hep-ph/9912329 [hep-ph]}
  \BibitemShut {NoStop}%
\bibitem [{\citenamefont {Henn}(2013)}]{Henn:2013pwa}%
  \BibitemOpen
  \bibfield  {author} {\bibinfo {author} {\bibfnamefont {J.~M.}\ \bibnamefont
  {Henn}},\ }\bibfield  {title} {\bibinfo {title} {{Multiloop integrals in
  dimensional regularization made simple}},\ }\href
  {https://doi.org/10.1103/PhysRevLett.110.251601} {\bibfield  {journal}
  {\bibinfo  {journal} {Phys. Rev. Lett.}\ }\textbf {\bibinfo {volume} {110}},\
  \bibinfo {pages} {251601} (\bibinfo {year} {2013})},\ \Eprint
  {https://arxiv.org/abs/1304.1806} {arXiv:1304.1806 [hep-th]} \BibitemShut
  {NoStop}%
\bibitem [{\citenamefont {Lee}(2015)}]{Lee:2014ioa}%
  \BibitemOpen
  \bibfield  {author} {\bibinfo {author} {\bibfnamefont {R.~N.}\ \bibnamefont
  {Lee}},\ }\bibfield  {title} {\bibinfo {title} {{Reducing differential
  equations for multiloop master integrals}},\ }\href
  {https://doi.org/10.1007/JHEP04(2015)108} {\bibfield  {journal} {\bibinfo
  {journal} {JHEP}\ }\textbf {\bibinfo {volume} {04}},\ \bibinfo {pages}
  {108}},\ \Eprint {https://arxiv.org/abs/1411.0911} {arXiv:1411.0911 [hep-ph]}
  \BibitemShut {NoStop}%
\bibitem [{\citenamefont {Chen}(1977)}]{Chen:1977oja}%
  \BibitemOpen
  \bibfield  {author} {\bibinfo {author} {\bibfnamefont {K.-T.}\ \bibnamefont
  {Chen}},\ }\bibfield  {title} {\bibinfo {title} {{Iterated path integrals}},\
  }\href {https://doi.org/10.1090/S0002-9904-1977-14320-6} {\bibfield
  {journal} {\bibinfo  {journal} {Bull. Am. Math. Soc.}\ }\textbf {\bibinfo
  {volume} {83}},\ \bibinfo {pages} {831} (\bibinfo {year} {1977})}\BibitemShut
  {NoStop}%
\bibitem [{\citenamefont {Henn}\ \emph
  {et~al.}(2020{\natexlab{a}})\citenamefont {Henn}, \citenamefont
  {Mistlberger}, \citenamefont {Smirnov},\ and\ \citenamefont
  {Wasser}}]{Henn:2020lye}%
  \BibitemOpen
  \bibfield  {author} {\bibinfo {author} {\bibfnamefont {J.}~\bibnamefont
  {Henn}}, \bibinfo {author} {\bibfnamefont {B.}~\bibnamefont {Mistlberger}},
  \bibinfo {author} {\bibfnamefont {V.~A.}\ \bibnamefont {Smirnov}},\ and\
  \bibinfo {author} {\bibfnamefont {P.}~\bibnamefont {Wasser}},\ }\bibfield
  {title} {\bibinfo {title} {{Constructing d-log integrands and computing
  master integrals for three-loop four-particle scattering}},\ }\href
  {https://doi.org/10.1007/JHEP04(2020)167} {\bibfield  {journal} {\bibinfo
  {journal} {JHEP}\ }\textbf {\bibinfo {volume} {04}},\ \bibinfo {pages}
  {167}},\ \Eprint {https://arxiv.org/abs/2002.09492} {arXiv:2002.09492
  [hep-ph]} \BibitemShut {NoStop}%
\bibitem [{\citenamefont {Dulat}\ and\ \citenamefont
  {Mistlberger}(2014)}]{Dulat:2014mda}%
  \BibitemOpen
  \bibfield  {author} {\bibinfo {author} {\bibfnamefont {F.}~\bibnamefont
  {Dulat}}\ and\ \bibinfo {author} {\bibfnamefont {B.}~\bibnamefont
  {Mistlberger}},\ }\bibfield  {title} {\bibinfo {title} {{Real-Virtual-Virtual
  contributions to the inclusive Higgs cross section at N3LO}},\ }\href@noop {}
  {\  (\bibinfo {year} {2014})},\ \Eprint {https://arxiv.org/abs/1411.3586}
  {arXiv:1411.3586 [hep-ph]} \BibitemShut {NoStop}%
\bibitem [{\citenamefont {Henn}\ and\ \citenamefont
  {Smirnov}(2013)}]{Henn:2013woa}%
  \BibitemOpen
  \bibfield  {author} {\bibinfo {author} {\bibfnamefont {J.~M.}\ \bibnamefont
  {Henn}}\ and\ \bibinfo {author} {\bibfnamefont {V.~A.}\ \bibnamefont
  {Smirnov}},\ }\bibfield  {title} {\bibinfo {title} {{Analytic results for
  two-loop master integrals for Bhabha scattering I}},\ }\href
  {https://doi.org/10.1007/JHEP11(2013)041} {\bibfield  {journal} {\bibinfo
  {journal} {JHEP}\ }\textbf {\bibinfo {volume} {11}},\ \bibinfo {pages}
  {041}},\ \Eprint {https://arxiv.org/abs/1307.4083} {arXiv:1307.4083 [hep-th]}
  \BibitemShut {NoStop}%
\bibitem [{\citenamefont {Almelid}\ \emph {et~al.}(2016)\citenamefont
  {Almelid}, \citenamefont {Duhr},\ and\ \citenamefont
  {Gardi}}]{Almelid:2015jia}%
  \BibitemOpen
  \bibfield  {author} {\bibinfo {author} {\bibfnamefont {O.}~\bibnamefont
  {Almelid}}, \bibinfo {author} {\bibfnamefont {C.}~\bibnamefont {Duhr}},\ and\
  \bibinfo {author} {\bibfnamefont {E.}~\bibnamefont {Gardi}},\ }\bibfield
  {title} {\bibinfo {title} {{Three-loop corrections to the soft anomalous
  dimension in multileg scattering}},\ }\href
  {https://doi.org/10.1103/PhysRevLett.117.172002} {\bibfield  {journal}
  {\bibinfo  {journal} {Phys. Rev. Lett.}\ }\textbf {\bibinfo {volume} {117}},\
  \bibinfo {pages} {172002} (\bibinfo {year} {2016})},\ \Eprint
  {https://arxiv.org/abs/1507.00047} {arXiv:1507.00047 [hep-ph]} \BibitemShut
  {NoStop}%
\bibitem [{\citenamefont {Aybat}\ \emph
  {et~al.}(2006{\natexlab{a}})\citenamefont {Aybat}, \citenamefont {Dixon},\
  and\ \citenamefont {Sterman}}]{Aybat:2006mz}%
  \BibitemOpen
  \bibfield  {author} {\bibinfo {author} {\bibfnamefont {S.~M.}\ \bibnamefont
  {Aybat}}, \bibinfo {author} {\bibfnamefont {L.~J.}\ \bibnamefont {Dixon}},\
  and\ \bibinfo {author} {\bibfnamefont {G.~F.}\ \bibnamefont {Sterman}},\
  }\bibfield  {title} {\bibinfo {title} {{The Two-loop soft anomalous dimension
  matrix and resummation at next-to-next-to leading pole}},\ }\href
  {https://doi.org/10.1103/PhysRevD.74.074004} {\bibfield  {journal} {\bibinfo
  {journal} {Phys. Rev. D}\ }\textbf {\bibinfo {volume} {74}},\ \bibinfo
  {pages} {074004} (\bibinfo {year} {2006}{\natexlab{a}})},\ \Eprint
  {https://arxiv.org/abs/hep-ph/0607309} {arXiv:hep-ph/0607309} \BibitemShut
  {NoStop}%
\bibitem [{\citenamefont {Aybat}\ \emph
  {et~al.}(2006{\natexlab{b}})\citenamefont {Aybat}, \citenamefont {Dixon},\
  and\ \citenamefont {Sterman}}]{Aybat:2006wq}%
  \BibitemOpen
  \bibfield  {author} {\bibinfo {author} {\bibfnamefont {S.~M.}\ \bibnamefont
  {Aybat}}, \bibinfo {author} {\bibfnamefont {L.~J.}\ \bibnamefont {Dixon}},\
  and\ \bibinfo {author} {\bibfnamefont {G.~F.}\ \bibnamefont {Sterman}},\
  }\bibfield  {title} {\bibinfo {title} {{The Two-loop anomalous dimension
  matrix for soft gluon exchange}},\ }\href
  {https://doi.org/10.1103/PhysRevLett.97.072001} {\bibfield  {journal}
  {\bibinfo  {journal} {Phys. Rev. Lett.}\ }\textbf {\bibinfo {volume} {97}},\
  \bibinfo {pages} {072001} (\bibinfo {year} {2006}{\natexlab{b}})},\ \Eprint
  {https://arxiv.org/abs/hep-ph/0606254} {arXiv:hep-ph/0606254} \BibitemShut
  {NoStop}%
\bibitem [{\citenamefont {Catani}(1998)}]{Catani:1998bh}%
  \BibitemOpen
  \bibfield  {author} {\bibinfo {author} {\bibfnamefont {S.}~\bibnamefont
  {Catani}},\ }\bibfield  {title} {\bibinfo {title} {{The Singular behavior of
  QCD amplitudes at two loop order}},\ }\href
  {https://doi.org/10.1016/S0370-2693(98)00332-3} {\bibfield  {journal}
  {\bibinfo  {journal} {Phys. Lett. B}\ }\textbf {\bibinfo {volume} {427}},\
  \bibinfo {pages} {161} (\bibinfo {year} {1998})},\ \Eprint
  {https://arxiv.org/abs/hep-ph/9802439} {arXiv:hep-ph/9802439} \BibitemShut
  {NoStop}%
\bibitem [{\citenamefont {Dixon}\ \emph {et~al.}(2008)\citenamefont {Dixon},
  \citenamefont {Magnea},\ and\ \citenamefont {Sterman}}]{Dixon:2008gr}%
  \BibitemOpen
  \bibfield  {author} {\bibinfo {author} {\bibfnamefont {L.~J.}\ \bibnamefont
  {Dixon}}, \bibinfo {author} {\bibfnamefont {L.}~\bibnamefont {Magnea}},\ and\
  \bibinfo {author} {\bibfnamefont {G.~F.}\ \bibnamefont {Sterman}},\
  }\bibfield  {title} {\bibinfo {title} {{Universal structure of subleading
  infrared poles in gauge theory amplitudes}},\ }\href
  {https://doi.org/10.1088/1126-6708/2008/08/022} {\bibfield  {journal}
  {\bibinfo  {journal} {JHEP}\ }\textbf {\bibinfo {volume} {08}},\ \bibinfo
  {pages} {022}},\ \Eprint {https://arxiv.org/abs/0805.3515} {arXiv:0805.3515
  [hep-ph]} \BibitemShut {NoStop}%
\bibitem [{\citenamefont {Korchemsky}\ and\ \citenamefont
  {Radyushkin}(1987)}]{Korchemsky:1987wg}%
  \BibitemOpen
  \bibfield  {author} {\bibinfo {author} {\bibfnamefont {G.~P.}\ \bibnamefont
  {Korchemsky}}\ and\ \bibinfo {author} {\bibfnamefont {A.~V.}\ \bibnamefont
  {Radyushkin}},\ }\bibfield  {title} {\bibinfo {title} {{Renormalization of
  the Wilson Loops Beyond the Leading Order}},\ }\href
  {https://doi.org/10.1016/0550-3213(87)90277-X} {\bibfield  {journal}
  {\bibinfo  {journal} {Nucl. Phys.}\ }\textbf {\bibinfo {volume} {B283}},\
  \bibinfo {pages} {342} (\bibinfo {year} {1987})}\BibitemShut {NoStop}%
\bibitem [{\citenamefont {Sterman}\ and\ \citenamefont
  {Tejeda-Yeomans}(2003)}]{Sterman:2002qn}%
  \BibitemOpen
  \bibfield  {author} {\bibinfo {author} {\bibfnamefont {G.~F.}\ \bibnamefont
  {Sterman}}\ and\ \bibinfo {author} {\bibfnamefont {M.~E.}\ \bibnamefont
  {Tejeda-Yeomans}},\ }\bibfield  {title} {\bibinfo {title} {{Multiloop
  amplitudes and resummation}},\ }\href
  {https://doi.org/10.1016/S0370-2693(02)03100-3} {\bibfield  {journal}
  {\bibinfo  {journal} {Phys. Lett. B}\ }\textbf {\bibinfo {volume} {552}},\
  \bibinfo {pages} {48} (\bibinfo {year} {2003})},\ \Eprint
  {https://arxiv.org/abs/hep-ph/0210130} {arXiv:hep-ph/0210130} \BibitemShut
  {NoStop}%
\bibitem [{\citenamefont {Becher}\ and\ \citenamefont
  {Neubert}(2020)}]{Becher:2019avh}%
  \BibitemOpen
  \bibfield  {author} {\bibinfo {author} {\bibfnamefont {T.}~\bibnamefont
  {Becher}}\ and\ \bibinfo {author} {\bibfnamefont {M.}~\bibnamefont
  {Neubert}},\ }\bibfield  {title} {\bibinfo {title} {{Infrared singularities
  of scattering amplitudes and N$^{3}$LL resummation for $n$-jet processes}},\
  }\href {https://doi.org/10.1007/JHEP01(2020)025} {\bibfield  {journal}
  {\bibinfo  {journal} {JHEP}\ }\textbf {\bibinfo {volume} {01}},\ \bibinfo
  {pages} {025}},\ \Eprint {https://arxiv.org/abs/1908.11379} {arXiv:1908.11379
  [hep-ph]} \BibitemShut {NoStop}%
\bibitem [{\citenamefont {Moch}\ \emph {et~al.}(2004)\citenamefont {Moch},
  \citenamefont {Vermaseren},\ and\ \citenamefont {Vogt}}]{Moch:2004pa}%
  \BibitemOpen
  \bibfield  {author} {\bibinfo {author} {\bibfnamefont {S.}~\bibnamefont
  {Moch}}, \bibinfo {author} {\bibfnamefont {J.~A.~M.}\ \bibnamefont
  {Vermaseren}},\ and\ \bibinfo {author} {\bibfnamefont {A.}~\bibnamefont
  {Vogt}},\ }\bibfield  {title} {\bibinfo {title} {{The Three loop splitting
  functions in QCD: The Nonsinglet case}},\ }\href
  {https://doi.org/10.1016/j.nuclphysb.2004.03.030} {\bibfield  {journal}
  {\bibinfo  {journal} {Nucl. Phys. B}\ }\textbf {\bibinfo {volume} {688}},\
  \bibinfo {pages} {101} (\bibinfo {year} {2004})},\ \Eprint
  {https://arxiv.org/abs/hep-ph/0403192} {arXiv:hep-ph/0403192} \BibitemShut
  {NoStop}%
\bibitem [{\citenamefont {Henn}\ \emph
  {et~al.}(2020{\natexlab{b}})\citenamefont {Henn}, \citenamefont
  {Korchemsky},\ and\ \citenamefont {Mistlberger}}]{Henn:2019swt}%
  \BibitemOpen
  \bibfield  {author} {\bibinfo {author} {\bibfnamefont {J.~M.}\ \bibnamefont
  {Henn}}, \bibinfo {author} {\bibfnamefont {G.~P.}\ \bibnamefont
  {Korchemsky}},\ and\ \bibinfo {author} {\bibfnamefont {B.}~\bibnamefont
  {Mistlberger}},\ }\bibfield  {title} {\bibinfo {title} {{The full four-loop
  cusp anomalous dimension in $\mathcal{N}=4$ super Yang-Mills and QCD}},\
  }\href {https://doi.org/10.1007/JHEP04(2020)018} {\bibfield  {journal}
  {\bibinfo  {journal} {JHEP}\ }\textbf {\bibinfo {volume} {04}},\ \bibinfo
  {pages} {018}},\ \Eprint {https://arxiv.org/abs/1911.10174} {arXiv:1911.10174
  [hep-th]} \BibitemShut {NoStop}%
\bibitem [{\citenamefont {von Manteuffel}\ \emph {et~al.}(2020)\citenamefont
  {von Manteuffel}, \citenamefont {Panzer},\ and\ \citenamefont
  {Schabinger}}]{vonManteuffel:2020vjv}%
  \BibitemOpen
  \bibfield  {author} {\bibinfo {author} {\bibfnamefont {A.}~\bibnamefont {von
  Manteuffel}}, \bibinfo {author} {\bibfnamefont {E.}~\bibnamefont {Panzer}},\
  and\ \bibinfo {author} {\bibfnamefont {R.~M.}\ \bibnamefont {Schabinger}},\
  }\bibfield  {title} {\bibinfo {title} {{Cusp and collinear anomalous
  dimensions in four-loop QCD from form factors}},\ }\href
  {https://doi.org/10.1103/PhysRevLett.124.162001} {\bibfield  {journal}
  {\bibinfo  {journal} {Phys. Rev. Lett.}\ }\textbf {\bibinfo {volume} {124}},\
  \bibinfo {pages} {162001} (\bibinfo {year} {2020})},\ \Eprint
  {https://arxiv.org/abs/2002.04617} {arXiv:2002.04617 [hep-ph]} \BibitemShut
  {NoStop}%
\bibitem [{\citenamefont {Agarwal}\ \emph {et~al.}(2021)\citenamefont
  {Agarwal}, \citenamefont {von Manteuffel}, \citenamefont {Panzer},\ and\
  \citenamefont {Schabinger}}]{Agarwal:2021zft}%
  \BibitemOpen
  \bibfield  {author} {\bibinfo {author} {\bibfnamefont {B.}~\bibnamefont
  {Agarwal}}, \bibinfo {author} {\bibfnamefont {A.}~\bibnamefont {von
  Manteuffel}}, \bibinfo {author} {\bibfnamefont {E.}~\bibnamefont {Panzer}},\
  and\ \bibinfo {author} {\bibfnamefont {R.~M.}\ \bibnamefont {Schabinger}},\
  }\bibfield  {title} {\bibinfo {title} {{Four-loop collinear anomalous
  dimensions in QCD and N=4 super Yang-Mills}},\ }\href
  {https://doi.org/10.1016/j.physletb.2021.136503} {\bibfield  {journal}
  {\bibinfo  {journal} {Phys. Lett. B}\ }\textbf {\bibinfo {volume} {820}},\
  \bibinfo {pages} {136503} (\bibinfo {year} {2021})},\ \Eprint
  {https://arxiv.org/abs/2102.09725} {arXiv:2102.09725 [hep-ph]} \BibitemShut
  {NoStop}%
\bibitem [{\citenamefont {Herzog}\ \emph {et~al.}(2023)\citenamefont {Herzog},
  \citenamefont {Ma}, \citenamefont {Mistlberger},\ and\ \citenamefont
  {Suresh}}]{Herzog:2023sgb}%
  \BibitemOpen
  \bibfield  {author} {\bibinfo {author} {\bibfnamefont {F.}~\bibnamefont
  {Herzog}}, \bibinfo {author} {\bibfnamefont {Y.}~\bibnamefont {Ma}}, \bibinfo
  {author} {\bibfnamefont {B.}~\bibnamefont {Mistlberger}},\ and\ \bibinfo
  {author} {\bibfnamefont {A.}~\bibnamefont {Suresh}},\ }\bibfield  {title}
  {\bibinfo {title} {{Single-soft emissions for amplitudes with two colored
  particles at three loops}},\ }\href {https://doi.org/10.1007/JHEP12(2023)023}
  {\bibfield  {journal} {\bibinfo  {journal} {JHEP}\ }\textbf {\bibinfo
  {volume} {12}},\ \bibinfo {pages} {023}},\ \Eprint
  {https://arxiv.org/abs/2309.07884} {arXiv:2309.07884 [hep-ph]} \BibitemShut
  {NoStop}%
\bibitem [{\citenamefont {Chen}\ \emph {et~al.}(2024)\citenamefont {Chen},
  \citenamefont {Luo}, \citenamefont {Yang},\ and\ \citenamefont
  {Zhu}}]{Chen:2023hmk}%
  \BibitemOpen
  \bibfield  {author} {\bibinfo {author} {\bibfnamefont {W.}~\bibnamefont
  {Chen}}, \bibinfo {author} {\bibfnamefont {M.-x.}\ \bibnamefont {Luo}},
  \bibinfo {author} {\bibfnamefont {T.-Z.}\ \bibnamefont {Yang}},\ and\
  \bibinfo {author} {\bibfnamefont {H.~X.}\ \bibnamefont {Zhu}},\ }\bibfield
  {title} {\bibinfo {title} {{Soft theorem to three loops in QCD and $
  \mathcal{N} $ = 4 super Yang-Mills theory}},\ }\href
  {https://doi.org/10.1007/JHEP01(2024)131} {\bibfield  {journal} {\bibinfo
  {journal} {JHEP}\ }\textbf {\bibinfo {volume} {01}},\ \bibinfo {pages}
  {131}},\ \Eprint {https://arxiv.org/abs/2309.03832} {arXiv:2309.03832
  [hep-ph]} \BibitemShut {NoStop}%
\bibitem [{\citenamefont {Guan}\ \emph
  {et~al.}(2025{\natexlab{b}})\citenamefont {Guan}, \citenamefont {Herzog},
  \citenamefont {Ma}, \citenamefont {Mistlberger},\ and\ \citenamefont
  {Suresh}}]{Guan:2024hlf}%
  \BibitemOpen
  \bibfield  {author} {\bibinfo {author} {\bibfnamefont {X.}~\bibnamefont
  {Guan}}, \bibinfo {author} {\bibfnamefont {F.}~\bibnamefont {Herzog}},
  \bibinfo {author} {\bibfnamefont {Y.}~\bibnamefont {Ma}}, \bibinfo {author}
  {\bibfnamefont {B.}~\bibnamefont {Mistlberger}},\ and\ \bibinfo {author}
  {\bibfnamefont {A.}~\bibnamefont {Suresh}},\ }\bibfield  {title} {\bibinfo
  {title} {{Splitting amplitudes at N$^{3}$LO in QCD}},\ }\href
  {https://doi.org/10.1007/JHEP01(2025)090} {\bibfield  {journal} {\bibinfo
  {journal} {JHEP}\ }\textbf {\bibinfo {volume} {01}},\ \bibinfo {pages}
  {090}},\ \Eprint {https://arxiv.org/abs/2408.03019} {arXiv:2408.03019
  [hep-ph]} \BibitemShut {NoStop}%
\bibitem [{\citenamefont {Brown}(2009)}]{Brown:2009qja}%
  \BibitemOpen
  \bibfield  {author} {\bibinfo {author} {\bibfnamefont {F.~C.~S.}\
  \bibnamefont {Brown}},\ }\bibfield  {title} {\bibinfo {title} {{Multiple zeta
  values and periods of moduli spaces M 0 ,n ( R )}},\ }\href@noop {}
  {\bibfield  {journal} {\bibinfo  {journal} {Annales Sci. Ecole Norm. Sup.}\
  }\textbf {\bibinfo {volume} {42}},\ \bibinfo {pages} {371} (\bibinfo {year}
  {2009})},\ \Eprint {https://arxiv.org/abs/math/0606419} {arXiv:math/0606419}
  \BibitemShut {NoStop}%
\bibitem [{\citenamefont {Goncharov}(2009)}]{Goncharov:2009lql}%
  \BibitemOpen
  \bibfield  {author} {\bibinfo {author} {\bibfnamefont {A.~B.}\ \bibnamefont
  {Goncharov}},\ }\bibfield  {title} {\bibinfo {title} {{A simple construction
  of Grassmannian polylogarithms}},\ }\href@noop {} {\  (\bibinfo {year}
  {2009})},\ \Eprint {https://arxiv.org/abs/0908.2238} {arXiv:0908.2238
  [math.AG]} \BibitemShut {NoStop}%
\bibitem [{\citenamefont {Chicherin}\ \emph {et~al.}(2020)\citenamefont
  {Chicherin}, \citenamefont {Henn},\ and\ \citenamefont
  {Papathanasiou}}]{Chicherin:2020umh}%
  \BibitemOpen
  \bibfield  {author} {\bibinfo {author} {\bibfnamefont {D.}~\bibnamefont
  {Chicherin}}, \bibinfo {author} {\bibfnamefont {J.~M.}\ \bibnamefont
  {Henn}},\ and\ \bibinfo {author} {\bibfnamefont {G.}~\bibnamefont
  {Papathanasiou}},\ }\bibfield  {title} {\bibinfo {title} {{Cluster algebras
  for Feynman integrals}},\ }\href@noop {} {\  (\bibinfo {year} {2020})},\
  \Eprint {https://arxiv.org/abs/2012.12285} {arXiv:2012.12285 [hep-th]}
  \BibitemShut {NoStop}%
\bibitem [{\citenamefont {Duhr}(2012)}]{Duhr:2012fh}%
  \BibitemOpen
  \bibfield  {author} {\bibinfo {author} {\bibfnamefont {C.}~\bibnamefont
  {Duhr}},\ }\bibfield  {title} {\bibinfo {title} {{Hopf algebras, coproducts
  and symbols: an application to Higgs boson amplitudes}},\ }\href
  {https://doi.org/10.1007/JHEP08(2012)043} {\bibfield  {journal} {\bibinfo
  {journal} {JHEP}\ }\textbf {\bibinfo {volume} {08}},\ \bibinfo {pages}
  {043}},\ \Eprint {https://arxiv.org/abs/1203.0454} {arXiv:1203.0454 [hep-ph]}
  \BibitemShut {NoStop}%
\bibitem [{\citenamefont {Brown}(2015)}]{Brown:2015ylf}%
  \BibitemOpen
  \bibfield  {author} {\bibinfo {author} {\bibfnamefont {F.}~\bibnamefont
  {Brown}},\ }\bibfield  {title} {\bibinfo {title} {{Notes on Motivic
  Periods}},\ }\href@noop {} {\  (\bibinfo {year} {2015})},\ \Eprint
  {https://arxiv.org/abs/1512.06410} {arXiv:1512.06410 [math.NT]} \BibitemShut
  {NoStop}%
\bibitem [{\citenamefont {Remiddi}\ and\ \citenamefont
  {Vermaseren}(2000)}]{Remiddi:1999ew}%
  \BibitemOpen
  \bibfield  {author} {\bibinfo {author} {\bibfnamefont {E.}~\bibnamefont
  {Remiddi}}\ and\ \bibinfo {author} {\bibfnamefont {J.~A.~M.}\ \bibnamefont
  {Vermaseren}},\ }\bibfield  {title} {\bibinfo {title} {{Harmonic
  polylogarithms}},\ }\href {https://doi.org/10.1142/S0217751X00000367}
  {\bibfield  {journal} {\bibinfo  {journal} {Int. J. Mod. Phys.}\ }\textbf
  {\bibinfo {volume} {A15}},\ \bibinfo {pages} {725} (\bibinfo {year}
  {2000})},\ \Eprint {https://arxiv.org/abs/hep-ph/9905237}
  {arXiv:hep-ph/9905237 [hep-ph]} \BibitemShut {NoStop}%
\bibitem [{\citenamefont {Schnetz}(2025)}]{Schnetz2025_HyperlogProcedures}%
  \BibitemOpen
  \bibfield  {author} {\bibinfo {author} {\bibfnamefont {O.}~\bibnamefont
  {Schnetz}},\ }\href
  {https://www2.mathematik.hu-berlin.de/~kreimer/tools/hyperlog\_procedures/}
  {\bibinfo {title} {Hyperlog procedures}} (\bibinfo {year} {2025}),\ \bibinfo
  {note} {maple package for single-valued hyperlogarithms and Feynman
  integrals}\BibitemShut {NoStop}%
\bibitem [{\citenamefont {Bauer}\ \emph {et~al.}(2000)\citenamefont {Bauer},
  \citenamefont {Kreckel},\ and\ \citenamefont {Frink}}]{Bauer2000}%
  \BibitemOpen
  \bibfield  {author} {\bibinfo {author} {\bibfnamefont {C.~W.}\ \bibnamefont
  {Bauer}}, \bibinfo {author} {\bibfnamefont {R.}~\bibnamefont {Kreckel}},\
  and\ \bibinfo {author} {\bibfnamefont {A.}~\bibnamefont {Frink}},\ }\bibfield
   {title} {\bibinfo {title} {{Introduction to the GiNaC framework for symbolic
  computation within the C++ programming language}},\ }\href@noop {} {\bibfield
   {journal} {\bibinfo  {journal} {J.Symb.Comput.}\ }\textbf {\bibinfo {volume}
  {33}},\ \bibinfo {pages} {1} (\bibinfo {year} {2000})}\BibitemShut {NoStop}%
\bibitem [{\citenamefont {Li}()}]{zhenjiempl}%
  \BibitemOpen
  \bibfield  {author} {\bibinfo {author} {\bibfnamefont {Z.-J.}\ \bibnamefont
  {Li}},\ }\href {https://github.com/munuxi/Multiple-Polylogarithm} {\bibinfo
  {journal} {https://github.com/munuxi/Multiple-Polylogarithm}\ }\BibitemShut
  {NoStop}%
\bibitem [{\citenamefont {Caron-Huot}\ \emph
  {et~al.}(2019{\natexlab{b}})\citenamefont {Caron-Huot}, \citenamefont
  {Dixon}, \citenamefont {Dulat}, \citenamefont {Von~Hippel}, \citenamefont
  {McLeod},\ and\ \citenamefont {Papathanasiou}}]{Caron-Huot:2019bsq}%
  \BibitemOpen
\bibfield  {journal} {  }\bibfield  {author} {\bibinfo {author} {\bibfnamefont
  {S.}~\bibnamefont {Caron-Huot}}, \bibinfo {author} {\bibfnamefont {L.~J.}\
  \bibnamefont {Dixon}}, \bibinfo {author} {\bibfnamefont {F.}~\bibnamefont
  {Dulat}}, \bibinfo {author} {\bibfnamefont {M.}~\bibnamefont {Von~Hippel}},
  \bibinfo {author} {\bibfnamefont {A.~J.}\ \bibnamefont {McLeod}},\ and\
  \bibinfo {author} {\bibfnamefont {G.}~\bibnamefont {Papathanasiou}},\
  }\bibfield  {title} {\bibinfo {title} {{The Cosmic Galois Group and Extended
  Steinmann Relations for Planar $\mathcal{N} = 4$ SYM Amplitudes}},\ }\href
  {https://doi.org/10.1007/JHEP09(2019)061} {\bibfield  {journal} {\bibinfo
  {journal} {JHEP}\ }\textbf {\bibinfo {volume} {09}},\ \bibinfo {pages}
  {061}},\ \Eprint {https://arxiv.org/abs/1906.07116} {arXiv:1906.07116
  [hep-th]} \BibitemShut {NoStop}%
\bibitem [{\citenamefont {Drummond}\ \emph {et~al.}(2018)\citenamefont
  {Drummond}, \citenamefont {Foster},\ and\ \citenamefont
  {G{\"u}rdo{\u{g}}an}}]{Drummond:2017ssj}%
  \BibitemOpen
  \bibfield  {author} {\bibinfo {author} {\bibfnamefont {J.}~\bibnamefont
  {Drummond}}, \bibinfo {author} {\bibfnamefont {J.}~\bibnamefont {Foster}},\
  and\ \bibinfo {author} {\bibfnamefont {{\"O}.}~\bibnamefont
  {G{\"u}rdo{\u{g}}an}},\ }\bibfield  {title} {\bibinfo {title} {{Cluster
  Adjacency Properties of Scattering Amplitudes in $N=4$ Supersymmetric
  Yang-Mills Theory}},\ }\href {https://doi.org/10.1103/PhysRevLett.120.161601}
  {\bibfield  {journal} {\bibinfo  {journal} {Phys. Rev. Lett.}\ }\textbf
  {\bibinfo {volume} {120}},\ \bibinfo {pages} {161601} (\bibinfo {year}
  {2018})},\ \Eprint {https://arxiv.org/abs/1710.10953} {arXiv:1710.10953
  [hep-th]} \BibitemShut {NoStop}%
\end{thebibliography}%

\end{document}